\documentclass[pra,aps,twocolumn,superscriptaddress,showpacs,nofootinbib]{revtex4-1}
\usepackage[utf8]{inputenc}
\usepackage{graphicx}
\usepackage{amsmath,amssymb,amsthm,mathrsfs,amsfonts,dsfont}
\usepackage{subfigure, epsfig}
\usepackage{braket}
\usepackage{bm}
\usepackage{enumerate}
\usepackage{framed}
\usepackage{tcolorbox}
\usepackage{multirow}
\usepackage{verbatim}
\usepackage{diagbox}

\newtheorem{theorem}{Theorem}
\newtheorem{lemma}{Lemma}

\newtheorem{definition}{Definition}

\newcommand{\Tr}{\mathrm{Tr}}

\newcommand{\Hmin}{H_{\text{min}}}

\newcommand{\Renyi}{\mathcal{R}_{\alpha}}

\newcommand{\f}[1]{f_{\mathrm{max}}#1}

\newcommand{\ketbra}[1]{\ket{#1}\bra{#1}}

\usepackage{url}

\usepackage{color}
\usepackage[10pt]{moresize}

\usepackage{amscd}

\usepackage{epsfig}
\usepackage{subfigure}

\makeatletter
\let\saved@includegraphics\includegraphics
\AtBeginDocument{\let\includegraphics\saved@includegraphics}
\renewenvironment*{figure}{\@float{figure}}{\end@float}

\makeatother

\begin{document}
\preprint{APS/123-QED}
\title{Experimental Realization of Device-Independent Quantum Randomness Expansion}
\author{Ming-Han Li}
\affiliation{Shanghai Branch, National Laboratory for Physical Sciences at Microscale and Department of Modern Physics, University of Science and Technology of China, Shanghai 201315, P.~R.~China}
\affiliation{Shanghai Branch, CAS Center for Excellence and Synergetic Innovation Center in Quantum Information and Quantum Physics, University of Science and Technology of China, Shanghai 201315, People's Republic of China}

\author{ Xingjian Zhang}
\affiliation{Center for Quantum Information, Institute for Interdisciplinary Information Sciences, Tsinghua University, Beijing 100084, People's Republic of China}

\author{Wen-Zhao Liu}
\author{Si-Ran Zhao}
\author{Bing Bai}
\author{Yang Liu}
\author{Qi Zhao}
\affiliation{Shanghai Branch, National Laboratory for Physical Sciences at Microscale and Department of Modern Physics, University of Science and Technology of China, Shanghai 201315, People's Republic of China}
\affiliation{Shanghai Branch, CAS Center for Excellence and Synergetic Innovation Center in Quantum Information and Quantum Physics, University of Science and Technology of China, Shanghai 201315, People's Republic of China}

\author{Yuxiang Peng}
\affiliation{Center for Quantum Information, Institute for Interdisciplinary Information Sciences, Tsinghua University, Beijing 100084, People's Republic of China}

\author{Jun Zhang}
\affiliation{Shanghai Branch, National Laboratory for Physical Sciences at Microscale and Department of Modern Physics, University of Science and Technology of China, Shanghai 201315, People's Republic of China}
\affiliation{Shanghai Branch, CAS Center for Excellence and Synergetic Innovation Center in Quantum Information and Quantum Physics, University of Science and Technology of China, Shanghai 201315, People's Republic of China}

\author{Yanbao Zhang}
\author{W. J. Munro}
\affiliation{NTT Basic Research Laboratories and NTT Research Center for Theoretical Quantum Physics, NTT Corporation, 3-1 Morinosato-Wakamiya, Atsugi, Kanagawa 243-0198, Japan}

\author{Xiongfeng Ma}
\affiliation{Center for Quantum Information, Institute for Interdisciplinary Information Sciences, Tsinghua University, Beijing 100084, P.~R.~China}

\author{Qiang Zhang}
\affiliation{Shanghai Branch, National Laboratory for Physical Sciences at Microscale and Department of Modern Physics, University of Science and Technology of China, Shanghai 201315, People's Republic of China}
\affiliation{Shanghai Branch, CAS Center for Excellence and Synergetic Innovation Center in Quantum Information and Quantum Physics, University of Science and Technology of China, Shanghai 201315, People's Republic of China}
\author{Jingyun Fan }
\affiliation{Shanghai Branch, National Laboratory for Physical Sciences at Microscale and Department of Modern Physics, University of Science and Technology of China, Shanghai 201315, People's Republic of China}
\affiliation{Shanghai Branch, CAS Center for Excellence and Synergetic Innovation Center in Quantum Information and Quantum Physics, University of Science and Technology of China, Shanghai 201315, People's Republic of China}
\affiliation{Shenzhen Institute for Quantum Science and Engineering and Department of Physics, Southern University of Science and Technology, Shenzhen, 518055, People's Republic of China}
\affiliation{Guangdong Provincial Key Laboratory of Quantum Science and Engineering, Southern University of Science and Technology, Shenzhen, 518055, People's Republic of China}
\author{Jian-Wei Pan}
\affiliation{Shanghai Branch, National Laboratory for Physical Sciences at Microscale and Department of Modern Physics, University of Science and Technology of China, Shanghai 201315, People's Republic of China}
\affiliation{Shanghai Branch, CAS Center for Excellence and Synergetic Innovation Center in Quantum Information and Quantum Physics, University of Science and Technology of China, Shanghai 201315, People's Republic of China}

\begin{abstract}
  Randomness expansion where one generates a longer sequence of random numbers from a short one is viable in quantum mechanics but not allowed classically. Device-independent quantum randomness expansion provides a randomness resource of the highest security level.
  Here we report the first experimental realization of device-independent quantum randomness expansion secure against quantum side information established through quantum probability estimation. We generate $5.47\times10^8$ quantum-proof random bits while consuming $4.39\times10^8$ bits of entropy, expanding our store of randomness by $1.08\times10^8$ bits
  at a latency of about 13.1 h, with a total soundness error $4.6\times10^{-10}$. Device-independent quantum randomness expansion not only enriches our understanding of randomness but also sets a solid base to bring quantum-certifiable random bits into realistic applications.
\end{abstract}

\maketitle

Randomness is a fundamental element of nature and ubiquitous in human activities.
Intrinsically, randomness comes from the breaking of quantum coherence~\cite{Ma2016QRNG,Acin_Certified_2016,RevModPhys.89.015004}.
The loophole-free violation of a Bell inequality~\cite{Bell, Hensen_Loophole_2015,Shalm15,Giustina15,Rosenfeld17,LiPRL2018} certifies entanglement, a special form of coherence, in a device-independent manner. This is the essence of device-independent quantum random number generation (DIQRNG)~\cite{Colbeck09,pironio2010,colbeck2011private},
and the rigorous security analysis makes it possible to design experiments secure against general attacks even under the extreme condition that the experimental devices themselves are not trusted~\cite{Ma2016QRNG,Acin_Certified_2016,RevModPhys.89.015004}.
The random bits certified in the loophole-free DIQRNG experiments are at the highest level of security among its kind being unpredictable to any strategies based on quantum or classical physics~\cite{liu2018device,bierhorst2018experimentally,Zhang2018Low}.
Randomness is required for setting the inputs of a Bell test, however, and in previous experimental realizations, more randomness is consumed than the certified ~\cite{pironio2010,liu2018device,Zhang2018Low,bierhorst2018experimentally}.
As it becomes publicly known after the experiment, the input randomness is consumed and cannot be reused. Otherwise, an adversary can take advantage of the information leakage and compromise the security of DIQRNG~\cite{Colbeck09}.
Theoretically, the amount of input randomness can be made arbitrarily small for the certification of the Bell inequality violation and further the randomness generation~\cite{Miller14}, and it is possible that the amount of generated randomness surpasses the input, which is randomness expansion.
Randomness expansion compensates the store of randomness for the consumption and provides more, eliminating the potential risk in security due to the \emph{circular} involvement of randomness.


The realization of device-independent quantum randomness expansion (DIQRE) has remained an outstanding challenge as it poses even stricter requirements than the loophole-free violation of Bell inequalities and DIQRNG.
In fact, the latter two tasks are prerequisites for randomness expansion. Besides, DIQRE requires a highly biased input probability distribution~\cite{Vazirani12,Miller14}, which causes larger statistical fluctuations and takes more statistics for successful certification.
Consequently, it is experimentally more demanding to realise DIQRE in a reasonable time. For instance, higher detection efficiency, higher visibility, and a more robust system behavior are required.
While entangled atomic systems~\cite{Hensen_Loophole_2015, Rosenfeld17} promise a large violation of Bell inequality, these systems are currently constrained by low event rates, making it hard to obtain decent experimental statistics within a reasonable time frame. Entangled photonic systems~\cite{Shalm15, Giustina15,LiPRL2018,Liu_High_2018,bierhorst2018experimentally,liu2018device,Zhang2018Low} on the other hand exhibit a relatively small violation of Bell inequality, but can be operated at very high event rates, thus providing an opportunity to achieve randomness expansion. We present here a concrete realization of DIQRE secure against a general quantum adversary\footnote{We do not consider the memory attack in reusing the protocol~\cite{barrett2013memory}. For one run of the experiment the memory attacks do not play a role (the randomness remains secure provided the devices used are kept secure after the experiment).} taking advantage of two recent advancements.
One is the development of cutting-edge single-photon detection with near unity efficiency~\cite{Eisaman2011}, which makes entangled photon-based loophole-free Bell test experiments viable. The other is the development of theoretical protocols~\cite{pironio2010,Vazirani12,Miller14,coudron2014infinite,arnon2018practical,knill2018quantum,brown2019framework}, which allows for the efficient generation of randomness secure against quantum side information in device-independent experiments, such as the quantum probability estimation (QPE) method~\cite{knill2018quantum}.
Below we briefly describe the spot-checking QPE method and our procedure to apply it to realize DIQRE.

\begin{figure}[tbh]
	\centering
	\resizebox{8cm}{!}{\includegraphics{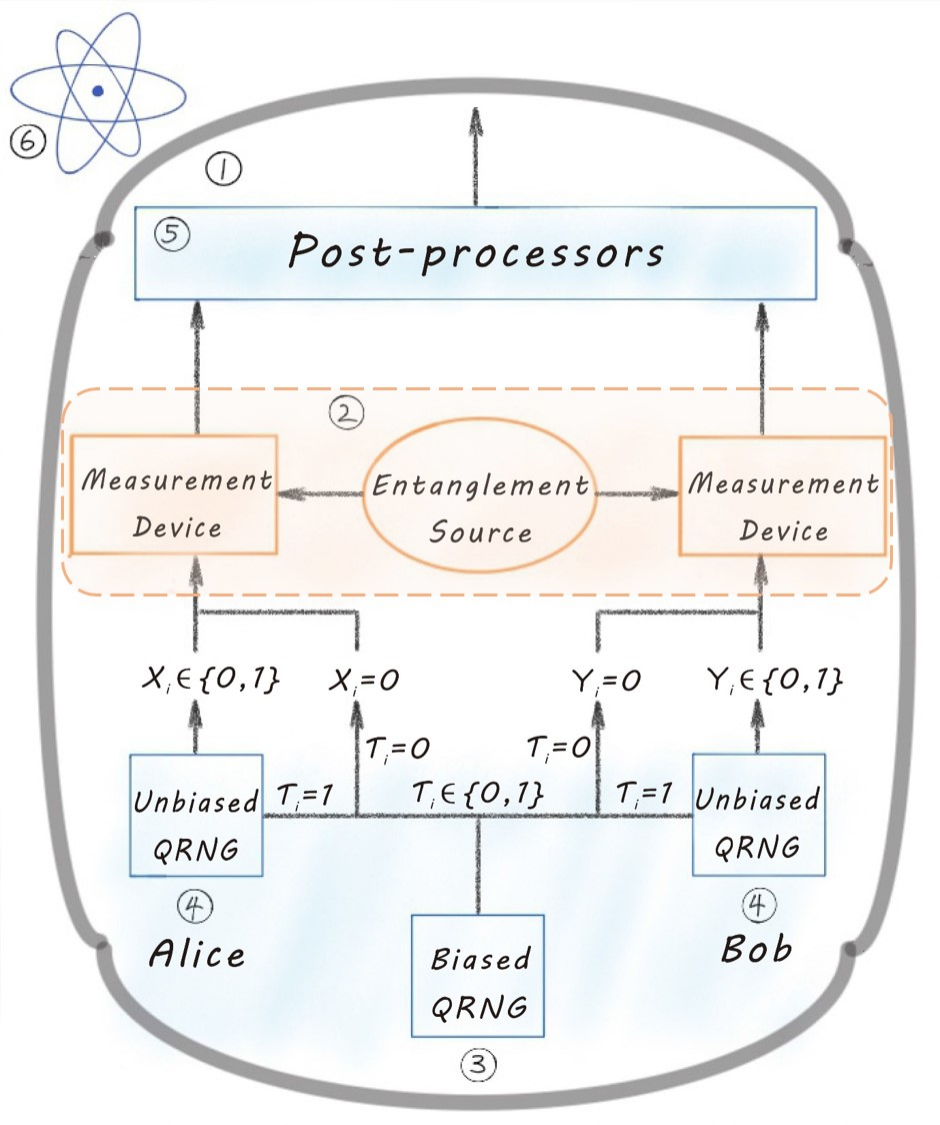}}
	
	\caption{
        A schematic demonstration of the experimental setup. The untrusted quantum devices are colored in orange and circled with the dotted line. The trusted parts are colored in blue. Specifically, we make the following assumptions in the protocol: (1) Secure lab: The information exchange with an outside entity is controlled.
  The devices cannot communicate to the outside to leak the experimental results directly.
(2) Nonsignaling condition: In each trial, the measurement process of Alice and Bob is independent of the other party.
(3) Trusted coordinator: A well characterized biased random number generator (depicted by ``Biased QRNG'' in the figure) determines a trial to be ``spot'' or ``checking.''
The setting is private to the measurement devices and the entanglement source.
(4) Trusted inputs: Alice and Bob each have a private random number generator (depicted by ``Unbiased QRNG'' in the figure) to feed perfect random bits to the measurement device in the ``checking trials.''
(5) Trusted postprocessors: The classical postprocessing procedure is trusted.
(6) Quantum mechanism: Quantum mechanics is correct and complete.
	}
	\label{Fig:Schematics}
\end{figure}

A procedure to realize randomness expansion according to the spot-checking QPE method is given in Box 1. The procedure consists of three key steps: parameter assignments, experimental randomness expansion and randomness extraction. The randomness expansion experiment is based on a sequence of Bell-test trials in the format of Clauser-Horne-Shimony-Holt (CHSH) game~\cite{CHSH} (see Fig.~\ref{Fig:Schematics} for experimental schematics and assumptions). In the $i$th trial, a source at the central station prepares a pair of entangled photons and sends them to two spatially separated parties, Alice and Bob. A coordinating random number generator independent of the measurement devices and the source generates a bit $T_i = 0$ (or $T_i = 1$) with nonzero probability $1-q$ (or $q$), respectively. If $T_i=0$, the trial is a ``spot trial'' where Alice and Bob set their input measurement settings to $X_i=0,\,Y_i=0$. If $T_i=1$, the trial is a ``checking trial'' where Alice and Bob choose their own measurement settings $X_i,\,Y_i\in\{0,1\}$ independently and uniformly at random. The value of $T_i$ is kept private from the measurement devices.
At the end of the trial Alice and Bob deliver an output $A_i,\,B_i\in\{0,1\}$, respectively. For a total number of $n$ experimental trials, we denote the input sequences by $\bm{X}=(X_1,X_2,\cdots,X_n),\,\bm{Y}=(Y_1,Y_2,\cdots,Y_n),\,\bm{T}=(T_1,T_2,\cdots,T_n)$, the outcome sequences by $\bm{A}=(A_1,A_2,\cdots,A_n),\,\bm{B}=(B_1,B_2,\cdots,B_n)$, respectively, and further denote $\bm{Z}=\bm{X}\bm{Y},\,\bm{C}=\bm{A}\bm{B}$ and $Z_i = X_iY_i,\,C_i = A_iB_i$.


In an adversarial picture, the source and measurement devices are prepared by a potentially malicious party, Eve, in advance of the experiment. Generally, the behavior of the quantum devices in each trial can be different and depend on previous events.
The final state after the experiment can be described by a classical-quantum state shared by Alice, Bob, and Eve,
$\rho_{\bm{CZ}E} = \sum_{\bm{c},\bm{z}}\ket{\bm{c}\bm{z}}\bra{\bm{c}\bm{z}}\otimes\rho_E(\bm{c}\bm{z})$.
We use lowercase letters to denote the values that the variables actually take in an experiment, and $(\bm{c},\bm{z})$ is one specific realization of the experiment's input-output sequence occurring with the probability $\Tr[\rho_E(\bm{c}\bm{z})]$.
The random variable $\bm{T}$ can be omitted without loss of generality, as we assume its value to be secret to Eve.
The quantum system of Eve, $\rho_E$, carries the quantum side information of the measurement results. We define the set of all possible joint final states after the experiment to be the model $\mathcal{M}(\bm{C},\bm{Z})$.

For randomness expansion, we determine a quantum estimation factor (QEF) $F(\bm{CZ})$ with power $\alpha$ for the model $\mathcal{M}(\bm{C},\bm{Z})$.
Informally speaking, with a fixed security parameter $\varepsilon_h$, the quantity $\{\log_2[F(\bm{cz})]+\log_2(\varepsilon_h^2/2)\}/(\alpha-1)$ witnesses the amount of private randomness extractable from the outputs $\bm{c}$.
To turn it into a rigorous statement, we need to determine a criterion of success before running the protocol.
The protocol succeeds if $F(\bm{cz})\geq2^{h_s(\alpha-1)}$, where $h_s$ (bits) is the success threshold. We denote the set of success events as $\Phi$ and $\kappa\in(0,1)$ as the predetermined lower bound on the success probability. Then, for an arbitrary state $\rho_{\bm{CZ}E}$ in the model, either the protocol success probability is less than $\kappa$, or the smooth min-entropy conditional on the success is lower bounded by~\cite{knill2018quantum}
\begin{equation}\label{Thm:QEFRandomness}
H_{\text{min}}^{\varepsilon_h}(\bm{C}|\bm{Z} E)_{\rho_{\bm{CZ}E|\Phi}}\geq h_s-\dfrac{1}{\alpha-1}\log_2\left(\dfrac{2}{\varepsilon_h^2}\right)+\dfrac{\alpha}{\alpha-1}\log_2\kappa,
\end{equation}
where $\varepsilon_h$ is the smoothing parameter in the smooth min-entropy of $\bm{C}$ conditioned on $\bm{Z}E$, and $\rho_{\bm{CZ}E|\Phi}$ represents the normalized state conditional on the success.
In particular, this result holds for the general condition.
To obtain a lower bound on the $\varepsilon_h$-smooth min-entropy, a lower bound $\kappa$ on the success probability is required.
Literature suggests setting $\kappa = \varepsilon_h$ to obtain a conservative lower bound on the $\varepsilon_h$-smooth min-entropy~\cite{arnon2018practical,knill2018quantum}.
However, we remark that the lower bound $\kappa$ is irrelevant for the soundness proof of the randomness generation protocol with QEFs (see Theorem 4 of Ref.~\cite{knill2018quantum}).
Because at each trial the probability distributions of the input variables $T_i$, $X_i$, and $Y_i$ are independent of the previous results and the quantum side information, a valid QEF $F(\bm{CZ})$ for a sequence of trials can be obtained by chaining the QEFs $F(C_iZ_i)$ for each experimental trial in the sequence (see Appendix~\ref{Supp:Quant_QEF}).

\begin{figure*}[hbt!]
\begin{tcolorbox}[title = {Box 1. Procedure for randomness expansion:}]
\begin{flushleft}
	{\bf Step 1. Parameter determination} \\(1) Assign the least target amount of entropy $k_{\text{exp}}$ (bits) to be expanded by;
	\\(2) Assign the soundness error $\varepsilon_S=2\varepsilon_h+\varepsilon_x\in(0,1)$ ($\varepsilon_h$ for randomness generation, $\varepsilon_x$ for randomness extraction);
	\\(3) Assign the probability distribution of $T_i$, $(1-q,q)$, with $0<q<1$.
	\\Based on these settings,\\
	(1) Determine a valid single trial QEF $F(CZ)$ with power $\alpha>1$;\\
	(2) Determine the success threshold for randomness expansion $h_s$ (bits);\\
	(3) Determine the largest allowed number of experimental trials $N$;\\
    (4) Determine the success probability of the protocol $\kappa\in(0,1)$.\\[0.5em]
	
	{\bf Step 2. Randomness expansion experiment} \\
(1) Before the experiment, set a classical register $G_0=1$. \\
(2) In the $i$th trial ($1\leq i\leq N$), \\
1) \emph{Measurements: }If $T_i=0$, set the measurement inputs as $X_i= Y_i=0$; if $T_i=1$, randomly set $X_i, Y_i\in\{0,1\}$. Record the measurement outputs $A_i,B_i$ and the corresponding QEF value $F_i(C_iZ_i)$.\\
2) \emph{Discrimination: }Update the register with $G_i = G_{i-1}F_i(C_i Z_i)$. If $\frac{1}{\alpha-1}\log_2G_i\geq h_s$, stop the experiment and set $F_j(c_j z_j)=1,j>n$. Goto Step 3. \\
(3) If $\frac{1}{\alpha-1}\log_2G_N< h_s$, abort the protocol.\\[0.5em]
	
	{\bf Step 3. Randomness extraction}: \\Apply a quantum-proof strong extractor to
$\bm{C}$ and obtain near-uniform random bits, with a security parameter no larger than $\varepsilon_x$.
\end{flushleft}
\end{tcolorbox}
\end{figure*}


If the success threshold is met in the experimental randomness expansion procedure, the protocol shall proceed to randomness extraction.
We use a quantum-proof strong extractor to extract certified random bits from the output sequence with a security parameter $\varepsilon_x$~\cite{Ma13} (see Appendix~\ref{extraction} for details in randomness extraction).
Informally speaking, the extractor takes the experimental output sequence $\bm{C}$ in the Bell test, together with a uniform bit string $S$, or the seed, as the input, and delivers a string of near-uniform random bits, except for a failure probability no larger than $\varepsilon_x$.
We do not consider the seed as entropy consumed in the experiment, since by definition the seed of a strong extractor can be reused albeit at the cost of the security parameter increased by $\varepsilon_x$~\cite{Ma13}.
Security is not compromised even if the seed is known by Eve after the execution of the protocol, as long as it is independent of the raw data and the classical postprocessing process is authenticated.
Guaranteed by the composable security property, the total failure probability of the protocol, or the soundness error, is no larger than $\varepsilon_S=2\varepsilon_h + \varepsilon_x$ (see Appendix~\ref{extraction}).

In the protocol, technically, an essential step is to find an (almost) optimal QEF for witnessing entropy in the experiment.
We begin with taking a sequence of experimental trials as the training set and use it to determine an empirical input-output probability distribution $\nu(CZ)$. With respect to this empirical probability distribution, we perform an optimization program to obtain a single trial QEF $F(CZ)$ and estimate the amount of output randomness per trial by $r_\nu(F,\alpha)= \mathbb{E}_{\nu}[\log_2F(CZ)]/(\alpha-1)$, without considering the smoothing parameter and protocol success probability.
For a robust experimental system whose behavior is near the empirical knowledge, the average output randomness witnessed by the QEF shall be close to $r_\nu(F,\alpha)$. All three QRNGs for the input settings are accounted for the input randomness, and the average entropy consumption per trial in randomness expansion is given by
\begin{equation}
  r_{\text{in}} = h(q)+2q,
\end{equation}
where $h(q) = -q\log_2q - (1-q)\log_2(1-q)$ is the binary entropy, and the coefficient $2$ is the amount of randomness consumed by Alice and Bob in a checking trial.
We would expect a successful randomness expansion if $r_{\nu}(F,\alpha)$ exceeds $r_{\text{in}}$.

Before the experiment, we fix the target least amount of near-uniform random bits to be expanded $k_{\text{exp}}$, the smoothing parameter $\varepsilon_h$ involved in randomness expansion,
the security parameter $\varepsilon_x$ in randomness extraction,
the success threshold $h_s$, the largest allowed number of experimental trials $N$,
and a lower bound on the success probability of the protocol $\kappa$ (see Appendix~\ref{sec:ParDet}). In the subsequent experiment, in each trial the single trial QEF takes a value $F(c_jz_j)$ with a realization $(c_jz_j)$, we keep updating a register $G_n$ by multiplying its value with the latest single-trial QEF value, where $n$ denotes the trial number. Before the experiment, the register $G_n$ is initialized to be $G_0=1$. We can stop the experiment in advance if the chained QEF value already exceeds the threshold before reaching the $N$th trial.

We realize randomness expansion on our upgraded photonic-entanglement distribution platform~\cite{Liu_High_2018,liu2018device,LiPRL2018}. A sketch of the experimental setup can be found in Appendix~\ref{Supp:Experiment}. In the experimental preparation, we enforce the nonsignaling condition by establishing spacelike separation between the measurement events of Alice and Bob, such that the output $A_i\,(B_i)$ is independent of $Y_i\,(X_i)$. We achieve a single-photon detection efficiency from creation to detection of $80.50\%$ for Alice and $82.20\%$ for Bob (see Appendix~\ref{efficiency}), and measure the average CHSH game value $J=0.750 88$, which surpasses the classical bound of $J\le0.75$ substantially over our previous results~\cite{Liu_High_2018,liu2018device,LiPRL2018}. Under this experimental condition, we determine a ratio of $1-q:q=8375:1\,(q=0.000 119)$ for a good randomness expansion result which corresponds to consuming the input entropy at a rate of $r_{\text{in}}=0.001 97$ (see Appendix~\ref{app:assign}). 
We operate our experiment with a 4 MHz repetition rate.

\begin{figure}[hbt!]
\centering
\resizebox{8cm}{!}{\includegraphics{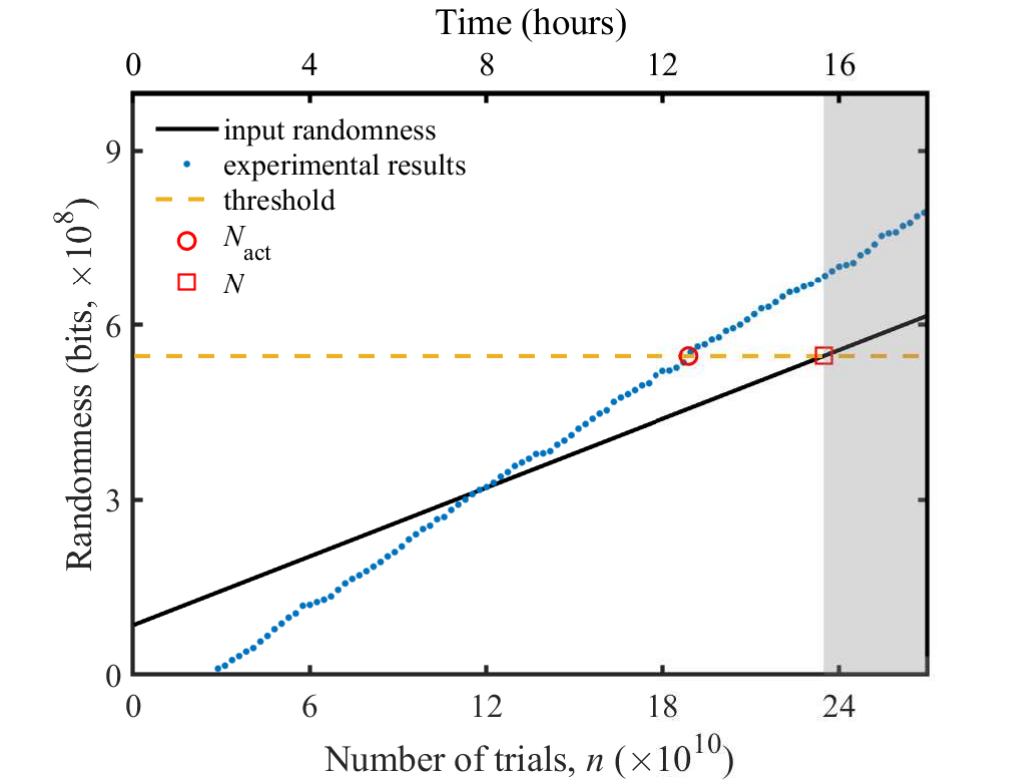}}
\caption{Randomness expansion of at least $k_{\text{exp}}=512$ random bits in at most $N=2.35\times10^{11}$ trials with a smoothing parameter $\varepsilon_h = 2^{-32}$ in randomness generation.
Blue dotted line: the experimental amount of generated $\varepsilon_h$-smooth min-entropy witnessed by the QEF by the $n$th trial $R_n$; black solid line: the amount of entropy consumed by the $n$th trial $k_0+nr_{\text{in}}$; yellow dash line: the least amount of $\varepsilon_h$-smooth min-entropy required to be generated for a successful randomness expansion task.
Surpassing the threshold before $N$ trials guarantees a successful randomness expansion of at least 512 near-uniform random bits at the end of the protocol, otherwise the protocol fails (represented by the gray area).
The open square denotes the largest allowed number of trials and the open circle denotes the actual number of trials to accomplish the task. We note that the experiment can be stopped in advance at the $N_{\text{act}}$th trial. Still, we continue the experiment for a longer period to show the robustness of the experimental setup, as shown in the shaded area (in all, we have accumulated data of approximately $3\times10^{12}$ trials). 
}
\label{Fig:512b}
\end{figure}

For this experimental demonstration, we set $k_{\text{exp}} = 512$ bits with a total soundness error of $\varepsilon_S = 2\varepsilon_h + \varepsilon_x \approx 2\times2^{-32}$ (with $\varepsilon_x = 2^{-100}$).
We conservatively set the bound on the protocol success probability $\kappa = \varepsilon_h$ in estimating the output randomness.
We take three hours training data (by consuming an amount of randomness $k_0 \approx 8.50\times10^7$ bits in $4.32\times 10^{10}$ trials), with which we determine a single trial QEF with power $\alpha = 1+1.172\times 10^{-6}$ and an expected output randomness rate $r_{\nu}(F,\alpha) = 0.002 89$ surpassing the input entropy rate by $0.000 92$. 
To determine the largest allowable number of trials,
we use the protocol success probability with honest devices $\gamma$, which relates to the completeness of the protocol (see Appendix~\ref{Supp:SecDef} for security definition). With $\gamma=99.3\%$, we determine the largest allowable number of trials to be $N = 2.35\times 10^{11}$ (open square in Fig.~\ref{Fig:512b}), which takes approximately 16 experimental hours, and set the threshold as $h_s=6.31\times10^8$ bits (see Appendix~\ref{512}). If $G_n$ surpasses $h_s$ in no more than $N$ trials, we shall expand our store of randomness by at least $512$ near-uniform random bits in the end.

Our result of the randomness expansion experiment (corresponding to Step 2 in Box 1) is shown in Fig.~\ref{Fig:512b}. We update $G_n$ with the observed QEF values for every latest accumulated 1-min of experimental data. To show the result of quantum expansion more directly, in Fig.~\ref{Fig:512b} we plot the quantity,
$R_n = [\log_2G_n-\log_2(2/\varepsilon_h^2) + \alpha\log_2\kappa]/(\alpha-1)$, as shown by the blue dotted line. In view of Eq.~\eqref{Thm:QEFRandomness} and the remark behind it, the quantity $R_n$ can be seen as the amount of $\varepsilon_h$-smooth min-entropy accumulated in the output after $n$ trials (\emph{if} the protocol will succeed).
Output randomness emerges after $2.6\times10^{10}$ trials and gradually surpasses the amount of entropy consumed $k_0+nr_{\text{in}}$ (black solid line). It surpasses the threshold (open circle) after $N_{\text{act}} = 1.89\times10^{11}$ trials (about 13.1 h).
At this point we can stop the experiment in advance and set the QEF values for the remained trials to be 1.
Therefore, we actually generate $5.47\times10^8$ bits of randomness and consume $k_0 + N_{\text{act}}r_{\text{in}} = 4.39\times10^8$ bits of entropy.
Afterward, we use the Toeplitz hashing matrix, a quantum-proof strong extractor~\cite{Ma13}, to extract near-uniform random bits with security parameter $\varepsilon_x=2^{-100}$.
In the end, we expand our store of randomness by $1.08\times10^8$ near-uniform random bits, which is far more than the requirement of the expansion task.

Our experimental realization of DIQRE is a substantial progress toward the ultimate understanding of randomness.
In particular, this may further inspire the research of other interesting directions of randomness, for example, randomness amplification~\cite{colbeck2012free}, which, instead of requiring input randomness to be independent of the devices, could amplify the imperfect random bits into perfect ones.
Besides, DIQRE might be related with quantum side information proof strong multisource extractors~\cite{kasher2010two}.
For these tasks, possible candidates for input randomness source could be cosmic randomness \cite{bell2004speakable} and human randomness \cite{BBT18}.
DIQRE, which expands a very small random seed to rather long sequence of random bits without compromising the security, possesses a great potential for realistic applications demanding high level secure randomness.

	The authors would like to thank X. Yuan and R. Colbeck for enlightening discussions, and E. Knill for pointing out a mistake in an earlier version. This work has been supported by the National Key R\&D Program of China (2017YFA0304000, 2017YFA0303900 and 2017YFA0304004), the National Fundamental Research Program (under Grant No.~2013CB336800), the National Natural Science Foundation of China, Chinese Academy of Science, Guangdong Provincial Key Laboratory Grant No.~2019B121203002, Guangdong Innovative and Entrepreneurial Research Team Program Grant No.~2019ZT08X324, the Key-Area Research and Development Program of Guangdong Province Grant No.~2020B0303010001, Shanghai Municipal Science and Technology Major Project (Grant No.~2019SHZDZX01), and the Zhongguancun Haihua Institute for Frontier Information Technology.

M.-H.~L and X.~Z. contributed equally to this work.

\emph{Note added.}---After finishing this work, we became aware of a similar DIQRE experiment work~\cite{shalm2019device}.
This work is based on the probability estimation method~\cite{zhang2018certifying}, which is closely related with the QPE method used in this work, but can only certify randomness in the presence of classical side information.
Besides, there is a DIQRE experiment against quantum side information which is based on the entropy accumulation theorem (EAT)~\cite{liu2019device}. We present a discussion on the comparison of QPE and EAT methods and list the related experimental results in Appendix~\ref{Supp:Comparison}.

\newpage
\onecolumngrid


\appendix

\section{Theory of Device-Independent Quantum Randomness Expansion}\label{theory}
In our randomness expansion experiment, we adopt the Clauser-Horne-Shimony-Holt-type (CHSH) Bell test in a spot-checking protocol.
The configuration involves three stations, commonly referred to as Alice, Bob and Charlie. In each trial, Charlie, whom Alice and Bob trust, draws a random bit $T\in\{0,1\}$ according to a Bernoulli distribution $(1-q,q)$. He then broadcasts the value of $T$ to Alice and Bob. If $T=0$, Alice and Bob both set their measurement inputs $X,Y$ as $0$. If $T=1$, Alice and Bob perform a standard CHSH-type Bell test. They make independent local measurements randomly, where Alice's setting choices form a random variable $X$ and Bob's form $Y$, $X,Y \in \{0,1\}$. Their outcomes form another two random variables $A,B$ ranging in the set $\{0,1\}$. We use subscripts to label the trial number, and letters without subscripts represent variables in a general single trial. Alice and Bob are space-like separated to block the signaling loophole in the Bell test.
When we refer to the devices as a whole for randomness expansion, we denote the input to the device as $Z=(X,Y)$ and output as $C=(A,B)$. Here we implicitly make the assumption that $T$ is kept private by Alice and Bob and not leaked to the outside during the process of randomness generation. $(C_i,Z_i)$ is called the result of the $i^{\text{th}}$ trial. For a sequence of trials, we use letters in bold, that is, $\bm{Z} = (Z_1,Z_2,\cdots) = (\bm{XY})$ and similar for other ones. Following convention, lowercase letters represent specific values the variables take in an experiment.

Ever since Colbeck first proposed the idea of quantum randomness expansion via Bell test in his PhD thesis~\cite{Colbeck09}, much progress has been made on the security analysis of device-independent quantum random number generation (DIQRNG)~\cite{pironio2010,Fehr13,Vazirani12,Miller14,miller2017universal}. So far there are two major protocols promising information-theoretically secure randomness generation in the presence of quantum side information and in a non-i.i.d. condition with current technology. One is based on entropy accumulation theorem (EAT), which requires a ``min-tradeoff function''~\cite{arnon2018practical}. A modified protocol further utilises NPA hierarchy method~\cite{brown2019framework}. The other protocol is the quantum probability estimation (QPE) theory based on quantum estimation factor (QEF)~\cite{knill2018quantum}, which directly estimates the output entropy from the observed statistics. As the QPE can be relatively more efficient for a small-size randomness generation and experiments with a small Bell violation, considering the characteristics of our optical platform and task target, we employ the QEF method in our device-independent quantum randomness expansion (DIQRE) experiment.

\subsection{Models in Quantum Probability Estimation Framework}\label{supp:model}
In QPE framework, the essence is a characterization of the set of possible final states in the experiment, which we call the model of the experiment.
In our device-independent test,
the possible final state of these systems can be described by a classical-quantum mixed state $\rho_{\bm{CZT}E}=\sum_{\bm{c},\bm{z},\bm{t}}\ket{\bm{c}\bm{z}\bm{t}}\bra{\bm{c}\bm{z}\bm{t}}\otimes\rho_E(\bm{c}\bm{z}\bm{t})$, where $\rho_E$ corresponds to the quantum side information of Eve.
According to our assumptions on the coordinating random number generator, Eve does not have access to $\bm{T}$. Hence we can omit the variable $\bm{T}$ in describing the set of final states, and describe the input $\bm{Z}$ by a probability distribution $\mu(\bm{Z})$. With the assumption that the choice of inputs is independent of the quantum systems, for $\mu(\bm{Z})$ belonging to a set of possible distributions $\mathcal{P}(\bm{Z})$, the set of possible final states is described by the model $\mathcal{M}(\bm{C},\bm{Z})$
\begin{equation}
\begin{aligned}
&\mathcal{M}(\bm{C},\bm{Z}) = \left\{\rho_{\bm{CZ}E}:\rho_{\bm{CZ}E}=\sum_{\bm{c},\bm{z}}\ket{\bm{c}\bm{z}}\bra{\bm{c}\bm{z}}\otimes\rho_E(\bm{c}\bm{z})\right\} =\mathcal{P}(\bm{Z})\mathcal{M}(\bm{C}|\bm{Z}),\\ &\textbf{Subject To:}\\
&\exists\rho_{ABE}\in\mathcal{D}(\mathcal{H}_{ABE}),\,\text{POVM Measurements } \{N_{\bm{x}}^{\bm{a}}\},\{N_{\bm{y}}^{\bm{b}}\},\,\mu(\bm{Z})\in\mathcal{P}(\bm{Z}), \\ &\text{such that } \forall \bm{c}=(\bm{a},\bm{b})\in\bm{C},\,\bm{z}=(\bm{x},\bm{y})\in\bm{Z},\\
&\rho_E(\bm{c}|\bm{z})=\Tr_{AB}[\rho_{ABE}(N_{\bm{x}}^{\bm{a}}\otimes N_{\bm{y}}^{\bm{b}}\otimes I_E)],\\
& \rho_E(\bm{cz})=\mu(\bm{z})\rho_E(\bm{c}|\bm{z}), \\
&\sum_{\bm{c}\in\bm{C}}\Tr[\rho_E(\bm{cz})]=\mu(\bm{z}).
\end{aligned}
\end{equation}
We denote the set of density operators acting on the space of Alice, Bob and Eve as $\mathcal{D}(\mathcal{H}_{ABE})$, and the positive operator-valued measure (POVM) measurements $\{N_{\bm{x}}^{\bm{a}}\},\{N_{\bm{y}}^{\bm{b}}\}$ act on the space of Alice and Bob, respectively, where the positive semi-definite operator $N_{\bm{x}}^{\bm{a}}$ denotes the operator corresponding to the input $\bm{x}$ and output $\bm{a}$ for Alice, and similarly for $N_{\bm{y}}^{\bm{b}}$ on Bob's side, and $\sum_{\bm{a}\in\bm{A}}N_{\bm{x}}^{\bm{a}}=I_A,\, \sum_{\bm{b}\in\bm{B}}N_{\bm{y}}^{\bm{b}}=I_B$.
$\mathcal{M}(\bm{C}|\bm{Z})$ is called the input-conditioned model related with $\mathcal{M}(\bm{C},\bm{Z})$, consisting of elements $\rho_{\bm{C}E|\bm{Z}}$, where for any $\bm{z}\in \bm{Z},\, \rho_{\bm{C}E|\bm{z}} = \dfrac{1}{\mu(\bm{z})}\sum_{\bm{c}}\ketbra{\bm{cz}}\otimes\rho_E(\bm{cz}) = \sum_{\bm{c}}\ketbra{\bm{cz}}\otimes\rho_E(\bm{c}|\bm{z})$, $\exists\rho_{\bm{CZ}E} = \sum_{\bm{c},\bm{z}}\ket{\bm{c}\bm{z}}\bra{\bm{c}\bm{z}}\otimes\rho_E(\bm{c}\bm{z}) \in\mathcal{M}(\bm{C},\bm{Z}),\, \mu(\bm{Z})\in\mathcal{P}(\bm{Z})$.
%
After the experiment, with a probability $\Tr[\rho_E(\bm{cz})]$ Alice and Bob observe the realisation $(\bm{c},\bm{z})$, and Eve's system becomes $\rho_E(\bm{cz})/\Tr[\rho_E(\bm{cz})]$.

In our experiment, the trials form an ordered sequence. Now let us consider the model for $n$ trials, which we now denote as $\mathcal{M}_{\leq n}(\bm{C}_{\leq n},\bm{Z}_{\leq n})$ to explicitly show the number of trials. If the result before the $n^{\text{th}}$ trial is $(\bm{c}_{<n},\bm{z}_{<n})$, we denote the model for the $n^{\text{th}}$ trial as $\mathcal{M}_{n}(C_{n},Z_{n})_{\bm{c}_{<n}\bm{z}_{<n}}$. For a variable $\rho_{\bm{CZ}E}\in\mathcal{M}_{\leq n}(\bm{C}_{\leq n},\bm{Z}_{\leq n})$, it should satisfy that for any $\bm{c}_{<n}\in\bm{C}_{<n},\,\bm{z}_{<n}\in\bm{Z}_{<n}$,
\begin{equation}\label{eq:ModelChain}
\begin{aligned}
    \sum_{\substack{c_{n}\in C_{n},\\z_{n}\in Z_{n}}}\rho_{\bm{CZ}E}&\in\mathcal{M}_{<n}(\bm{C}_{<n},\bm{Z}_{<n}),\\ \rho_{\bm{c}_{<n}\bm{z}_{<n}C_nZ_nE}&\in\mathcal{M}_{n}(C_{n},Z_{n})_{\bm{c}_{<n}\bm{z}_{<n}},
\end{aligned}
\end{equation}
where
\begin{equation}
  \rho_{\bm{c}_{<n}\bm{z}_{<n}C_nZ_nE} = \sum_{c_n,z_n}\ketbra{\bm{c}_{<n}\bm{z}_{<n}c_nz_n}\otimes\rho_E(c_nz_n).
\end{equation}
Apart from the above constraint, we additionally have the Markov chain condition between adjacent trials, $\bm{C}_{<i}\leftrightarrow\bm{Z}_{<i}E\leftrightarrow Z_i,\,\forall i$, guaranteed by the assumption of the independent and identically distributed (i.i.d.) input settings in our experiment. The quantum Markov chain is defined as
\begin{definition}(Quantum Markov Chain)
For a tri-partite quantum state $\rho_{ABC}\in\mathcal{D}(\mathcal{H}_{ABC})$, it is said to obey a quantum Markov chain condition $A\leftrightarrow B\leftrightarrow C$, if for $\mathcal{H}_{B}$, there exists a decomposition $\mathcal{H}_{B}=\bigoplus_{\lambda}\mathcal{H}_{a_\lambda}\otimes\mathcal{H}_{c_\lambda}$, such that
\begin{equation}
    \rho_{ABC} = \bigoplus_{\lambda}p_{\lambda}\tau_{Aa_\lambda}\otimes\chi_{c_\lambda C},\,\tau_{Aa_\lambda}\in\mathcal{D}(\mathcal{H}_{A}\otimes\mathcal{H}_{a_\lambda}),\,\chi_{c_\lambda C}\in\mathcal{D}(\mathcal{H}_{c_\lambda}\otimes\mathcal{H}_{C}),\,p_{\lambda}\geq0,\,\forall\lambda.
\end{equation}
\end{definition}
With Eq.~\eqref{eq:ModelChain} and the Markov chain condition, the model in our experiment $\mathcal{C}(\bm{C},\bm{Z})$ can be chained by the models for each single round (see Sec.~VIB in~\cite{knill2018quantum}),
\begin{equation}
  \mathcal{M}_{\leq n}(\bm{C}_{\leq n},\bm{Z}_{\leq n}) = \mathcal{M}_{<n}(\bm{C}_{<n},\bm{Z}_{<n})\circ \mathcal{M}_{n}(C_{n},Z_{n})_{\bm{c}_{<n}\bm{z}_{<n}} = \mathcal{M}_{1}(C_{1},Z_{1}) \circ \mathcal{M}_{2}(C_{2},Z_{2})_{c_1z_1} \circ \cdots \mathcal{M}_{n}(C_{n},Z_{n})_{\bm{c}_{<n}\bm{z}_{<n}}.
\end{equation}
Therefore, we only need to focus on the model for a single round of the experiment.
Suppose the initial state before measurements shared by Alice, Bob and Eve is $\rho_{ABE}$, and the result observed by Alice and Bob is $z = (x,y),\,c = (a,b)$. The POVM element can be expressed as $P_{cz} = Q_{a|x}\otimes Q_{b|y}$, with $Q_{a|x}, Q_{b|y}\succeq 0,\,\sum_{a}Q_{a|x} = \sum_{b}Q_{b|y} = \mathbb{I},\,\forall a,b,x,y$. If the input $Z$ is drawn according to a probability distribution $\mu(Z)$, the final state is
\begin{equation}
    \rho_{CZE} = \sum_{c,z}\ket{cz}\bra{cz}\otimes\left(\mu(z)\Tr_{AB}[\rho_{ABE}(P_{c|z}\otimes\mathbb{I}_E)]\right).
\end{equation}
%
The model for a single trial of CHSH-type Bell test has been well investigated~\cite{pironio2009device}. Here we use the results of Theorem 6 in~\cite{knill2018quantum} for subsequent analysis.
For a single trial, under a fixed input probability distribution $\mu(Z)$, the model can be expressed as the convex combination of states in the form
\begin{equation}\label{eq:CHSHModel}
    \rho_{CZE} = \sum_{c,z}\mu(z)\ketbra{cz}\otimes U\tau^{1/2}P_{c|z}\tau^{1/2}U^{\dag},
\end{equation}
where $\tau\succeq 0, U$ is an isometry: $(\mathbb{C}^2)^{\otimes 2}\rightarrow\mathcal{H}_E$. With a discussion on the dimension, it is known that $\mathcal{H}_E$ can be restricted to $\mathbb{C}^4$~\cite{Tsirelson}. The measurement operator $P_{C|Z;\theta} = Q_{A|X;\theta_1}\otimes Q_{B|Y;\theta_2}$. Here we introduce the parameter $\theta = (\theta_1,\theta_2)$ to characterize the measurement operators, where
\begin{equation}\label{eq:CHSHMeasure}
\begin{aligned}
    &Q_{a|0;\theta_1}=\frac{\mathbb{I}+(-1)^{a}\sigma_z}{2}, \\
    &Q_{a|1;\theta_1}=\frac{\mathbb{I}+(-1)^{a}[\cos(\theta_1)\sigma_z+\sin(\theta_1)\sigma_x]}{2}, \,\theta_1\in (-\pi,\pi],
\end{aligned}
\end{equation}
and a similar representation holds for Bob's measurement operators. Combining Eq.~\eqref{eq:CHSHModel} and~\eqref{eq:CHSHMeasure}, we obtain the model for the single round $\mathcal{M}(C,Z)$ and its corresponding input-conditioned model $\mathcal{M}(C|Z)$ under a fixed input probability distribution $\mu(Z)$.

If we take into consideration the possible bias in the input probability distribution, when the actual probability distribution $\tilde{\mu}(Z)$ can be expressed as a convex combination of several extremal probability distributions $\{\mu_i(Z)\}_i$, we can denote the set of input probability distributions as the convex span of these probability distributions, $\text{span}\{\mu_i(Z)\}$, and the single trial model is
\begin{equation}
\mathcal{M}(C,Z)=\text{span}\{\mu_i(Z)\}\times\mathcal{M}(C|Z) = \{\tilde{\mu}(Z)\rho_{CE|Z}:\rho_{CE|Z}\in\mathcal{M}(C|Z),\,\tilde{\mu}(Z)\in\text{span}\{\mu_i(Z)\}\}.
\end{equation}

\subsection{Quantification of Randomness with Quantum Estimation Factors}\label{Supp:Quant_QEF}
For a concrete model $\mathcal{M}(\bm{C},\bm{Z})$, we can define quantum estimation factors (QEF), which is related to $\alpha$-R\'enyi powers.
\begin{definition}
    (R\'enyi Powers) Let $\rho\succeq 0$, and the support of $\rho$ lies in $\sigma\succeq 0$, $\beta = \alpha-1 >0$. The R\'enyi power of order $\alpha$ of $\rho$ conditional on $\sigma$ is
    \begin{equation}
      \Renyi(\rho|\sigma) = \Tr\left[\left(\sigma^{-\beta/(2\alpha)}\rho\sigma^{-\beta/(2\alpha)}\right)^\alpha\right].
    \end{equation}
\end{definition}

\begin{definition}
    (Quantum Estimation Factor) The non-negative real-valued function $F(\bm{CZ})$ is a quantum estimation factor (QEF) with power $\alpha>1$ for the model $\mathcal{M}(\bm{C},\bm{Z})$, if $F(\bm{CZ})$ satisfies the following inequality with $\alpha$-R\'enyi power for $\forall\tau\in\mathcal{M}(\bm{C},\bm{Z})$
    \begin{equation}
    \begin{aligned}
      &\sum_{\bm{c,z}}F(\bm{cz})\mathcal{R}_{\alpha}(\tau_E(\bm{cz})|\tau_E(\bm{z}))\leq 1,\,\alpha > 1,\\
      &\tau = \sum_{\bm{c},\bm{z}}\ket{\bm{c}\bm{z}}\bra{\bm{c}\bm{z}}\otimes\tau_E(\bm{c}\bm{z}),\,\tau_E(\bm{z}) = \sum_{\bm{c}}\tau_E(\bm{cz}).
    \end{aligned}
    \end{equation}
\end{definition}

In the following we write $\mathcal{R}_{\alpha}(\tau(\bm{cz})|\tau(\bm{z}))\equiv\Renyi(\bm{cz}|\bm{z})_{\tau(\bm{cz})}$ for brevity and in accordance with the main text. With QEFs, we can give a lower bound to the amount of randomness that can be extracted from the device-independent (DI) test, which is measured by the smooth conditional min-entropy~\cite{Tomamichel2010duality}.
\begin{definition}(Smooth Conditional Min-Entropy)
  Consider a quantum state $\rho\in\mathcal{D}(\mathcal{H}_{CZE})$. The $\varepsilon_h$-smooth min-entropy of $C$ conditioned on $Z,E$ is
  \begin{equation}
  \begin{aligned}
    &\Hmin^{\varepsilon_h}(\bm{C}|\bm{Z}E)_{\rho} = \max_{\substack{P(\rho',\rho)\leq\varepsilon_h,\\ \rho'\in\mathcal{S}(\mathcal{H}_{CZE})}} \Hmin(\bm{C}|\bm{Z}E)_{\rho'},\\
    &\Hmin(\bm{C}|\bm{Z}E)_{\rho'} = \sup_{\sigma\in\mathcal{S}(\mathcal{H}_{ZE})}\sup_{\lambda}\{\lambda\in\mathbb{R}: \rho'\leq\exp(-\lambda)I_C\otimes\sigma\},
  \end{aligned}
  \end{equation}
where $P(\rho',\rho)$ is the purified distance between $\rho',\rho$, and $\mathcal{S}(\cdot)$ denotes the set of sub-normalised density operators acting on the corresponding Hilbert space. The purified distance between $\rho,\tau\in\mathcal{S}(\mathcal{H})$ is defined as
\begin{equation}
\begin{aligned}
  P(\rho,\tau)=\sqrt{1-\left(\Tr|\sqrt{\rho}\sqrt{\tau}|+\sqrt{(1-\Tr[\rho])(1-\Tr[\tau])}\right)^2}
\end{aligned}
\end{equation}
\end{definition}

\begin{theorem}(Theorem 3 in~\cite{knill2018quantum})
Prior to a randomness generation procedure with $N$ trials, suppose that $F(\bm{CZ})$ is a QEF with power $\alpha$ for the model of the experiment $\mathcal{M}(\bm{C},\bm{Z})$, and set the smoothing parameter in randomness generation $\varepsilon_h\in(0,1]$, the threshold $h_s>0$ for a successful randomness generation and a lower bound $\kappa\in(0,1]$ to the probability of a successful randomness generation.
	A successful randomness generation is that, after $N$ trials have been executed, an event $(\bm{c},\bm{z})$ happens such that $F(\bm{cz})\geq 2^{h_s(\alpha-1)}$. In this case, for any possible final state $\rho\in\mathcal{M}(\bm{C},\bm{Z})$, we have \emph{either} the probability of success is less than $\kappa$, \emph{or} the quantum smooth conditional min-entropy given success satisfies
	\begin{equation}\label{eq:QEFRandomness}
	\Hmin^{\varepsilon_h}(\bm{C}|\bm{Z} E)_{\rho}\geq h_s-\dfrac{1}{\alpha-1}\log_2\left(\dfrac{2}{\varepsilon_h^2}\right)+\dfrac{\alpha}{\alpha-1}\log_2\kappa.
	\end{equation}
\end{theorem}

As we have the quantum Markov condition in our protocol, $\bm{C}_{<i}\leftrightarrow \bm{Z}_{<i}E\leftrightarrow Z_i,\,\forall i$, 
we have the following result:
\begin{theorem}[Theorem 5 in~\cite{knill2018quantum}]\label{Thm:chainQEF}
	Under the condition $\bm{C}_{<i}\leftrightarrow \bm{Z}_{<i}E\leftrightarrow Z_i,\,\forall i$, let $G_n(\bm{CZ})$ be a QEF with power $\alpha$ for the model $\mathcal{M}_{\leq n}(\bm{C},\bm{Z})$ of the first $n$ trials, and $F_{n+1}(C_{n+1},Z_{n+1})_{\bm{cz}}$ be a QEF with the same power for the model $\mathcal{M}_{n+1}(C_{n+1},Z_{n+1})_{\bm{cz}}$ of the $(n+1)^\text{th}$ trial conditioned on the result $\bm{c},\bm{z}$ of previous experimental trials. Then $G_n(\bm{CZ})\cdot F_{n+1}(CZ)_{\bm{CZ}}$ is a valid QEF with power $\alpha$ for the model $\mathcal{M}_{\leq n+1}(\bm{C},\bm{Z})$ of the first $(n+1)$ trials.
\end{theorem}
Therefore, we need only consider the QEF for the model of a single round.

\subsection{Optimization of Quantum Estimation Factor}\label{supp:opt_qef_sec}
In a DIQRE experiment, we need to optimize the QEF used for witnessing quantum randomness. If the experiment has a stable behaviour, we may expect the existence of some probability distribution $\nu(CZ)$ behind the result, and it is adapted to the model for a single trial, i.e. $\exists\rho_{CZE}\in\mathcal{M}(C,Z),\,\rho_{CZE}=\sum_{c,z}\ket{cz}\bra{cz}\otimes\rho_E(cz), \,\nu(cz)=\Tr[\rho_E(cz)],\,\forall c\in C,\,z\in Z$.
Thanks to Theorem~\ref{Thm:chainQEF}, we can chain single trial QEFs to derive a valid QEF for multiple trials, as long as the Markov chain condition is guaranteed between adjacent trials. As the model for each trial is included in $\mathcal{M}(C,Z)$, we can use the same QEF for all trials before the stopping criterion is met, and optimize it to maximize an expected randomness rate witnessed by QEF


\begin{equation}\label{QEFopt}
\begin{aligned}
    &\max_{F(CZ),\,\alpha} r_{\nu}(F,\alpha) = \frac{1}{\alpha-1}\sum_{cz}\nu(cz)\log_2(F(cz)),\\ &\text{s.t.} \left\{
        \begin{array}{lr}
          F(cz)\geq 0,\,\forall c\in C,\,z\in Z, &\\
          \sum_{cz}F(cz)\Renyi(cz|z)_{\tau_E(cz)}\leq 1,\,\forall\tau\in\mathcal{M}(C,Z), &\\
          \alpha>1.
        \end{array}
          \right.
\end{aligned}
\end{equation}

No efficient direct optimization of Eq.~\eqref{QEFopt} has been developed, though. While a sub-optimal solution can be accepted, as long as it does not compromise security, i.e. the solution satisfies the constraint in Eq.~\eqref{QEFopt}. A heuristic solution is to first solve a similar optimization, however, wherein the adversary is constrained to classical side information, and scale the solution with some coefficient to derive a valid QEF. The first step of the optimization is the so-called optimization of probability estimation factor (PEF), proposed in~\cite{zhang2018certifying}. We write the PEF as $F'(CZ)$. In this case Eve's system is restricted to be one-dimensional. We write the corresponding model as $\mathcal{M}_C(C,Z)$, which becomes a set of joint probability distribution $\tau'(CZ)$, and
the optimization problem can be taken as
\begin{equation}\label{PEFopt}
 \begin{aligned}
    &\max_{F'(CZ),\alpha} r_{\nu}(F',\alpha) = \frac{1}{\alpha-1}\sum_{cz}\nu(cz)\log_2(F'(cz)),\\
    &\text{s.t.} \left\{
        \begin{array}{lr}
          F'(CZ)\geq 0,\,\forall c\in C,\,z\in Z, &\\
          \sum_{cz}F'(cz)\tau'(c|z)^{\alpha-1}\tau'(cz)\leq 1,\,\forall\tau'(CZ)\in\mathcal{M}_C(C,Z), &\\
          \alpha>1.
        \end{array}
          \right.
 \end{aligned}
\end{equation}
This optimization problem is a concave maximization problem over a convex set, to which a global optimal solution can be found in principle.
Directly solving this optimization problem is still difficult, though. The probability distributions come from measuring quantum states. Possible probability distributions form a convex set, yet not a polytope, hence going over the solving domain determined by the constraints is no easy task. While suboptimal solutions to this method can be derived with an appropriate extension of the solving domain into a polytope.
In our experiment, we optimize the PEF over a polytope determined by 8 PR-boxes~\cite{prbox} and the corresponding Tsirelson's bounds~\cite{Tsirelson}, which include all input-conditional distributions of outputs according to quantum mechanics. In all, the modified optimization problem is defined over a polytope with 80 extreme points, and they can be constructed by PR-boxes and local deterministic points~\cite{knill2017quantum, Bierhorst16secrecy}. Then the optimization becomes
\begin{equation}\label{PEFoptTsirelson}
 \begin{aligned}
    &\max_{F'(CZ),\alpha} r_{\nu}(F',\alpha) = \frac{1}{\alpha-1}\sum_{cz}\nu(cz)\log_2(F'(cz)),\\
    &\text{s.t.} \left\{
        \begin{array}{lr}
          F'(CZ)> 0,\,\forall c\in C,\,z\in Z, &\\
          \sum_{cz}F'(cz)\tau'_{k}(c|z)^{\alpha-1}\tau'_{k}(cz)\leq 1,\,k=1,2,\cdots,80, &\\
          \alpha>1.
        \end{array}
          \right.
 \end{aligned}
\end{equation}
$\tau'_k(CZ)$ are the extremal points of this convex polytope model.

It needs to be noted that $\nu(CZ)$ should be adapted to the model. In our experiment, we derive this empirical probability distribution from actual experimental data. The raw conditional frequency distribution $p(C|Z)$ attained from the experiment has a weak signaling behaviour between Alice and Bob due to the statistical fluctuation. For this matter, we conduct  maximum-likelihood estimation  to derive a maximum likely probability distribution with respect to the raw frequency distribution and adapted to the non-signaling condition,
\begin{equation}\label{MLE}
 \begin{aligned}
    &\max_{\nu} \sum_{abxy}p(ab|xy)\log\nu(abxy),\\
    &\text{s.t. }\left\{
        \begin{array}{lr}
          \nu(xy)=\mu(xy), &\\
          \nu(a|xy)=\nu(a|x), \nu(b|xy)=\nu(b|y).
        \end{array}
          \right.
 \end{aligned}
\end{equation}
Here we write the variables $Z=(X,Y),\,C=(A,B)$ explicitly. We fix the input distribution to be $\mu(XY)$, which is the ideal input probability distribution.

After obtaining the optimal PEF $F'(CZ)$, we scale it by a parameter $f_{\text{max}}$ to obtain a QEF. We first normalize $F'(CZ)$ with $f_0$ such that $\frac{1}{f_0}\sum_{cz}F'(cz)\equiv\sum_{cz}\tilde{F}'(cz)=1$. Then we introduce a parameter $\tilde{f}$ and solve the optimization problem
\begin{equation}\label{fmax}
 \begin{aligned}
    &\tilde{f}=\max_{\tau,\theta}\sum_{cz}\mu(z)\tilde{F}'(cz)(\mathrm{Tr}[P_{c|z;\theta}\tau^{1/{\alpha}}P_{c|z;\theta}])^{\alpha}, \\
    &\text{s.t. }\theta=(\theta_1,\theta_2)\in[0,\pi]\times[0,\pi],\,\tau\succeq 0 \text{ with }\Tr[\tau]=1.
 \end{aligned}
\end{equation}
The value $\tilde{F}'(CZ)/\tilde{f}\equiv F'(CZ)/\f$ delivers a valid QEF. We denote $\tilde{f}=\max_\tau\tilde{f}(\theta)$, where
\begin{equation}
  \tilde{f}(\theta)=\max_{\tau}\sum_{cz}\mu(z)\tilde{F}'(cz)(\mathrm{Tr}[P_{c|z;\theta}\tau^{1/{\alpha}}P_{c|z;\theta}])^{\alpha}.
\end{equation}
For a fixed tuple $\theta=(\theta_1,\theta_2)$, $\tilde{f}(\theta)$ is concave with respect to $\tau$, and we apply a Frank-Wolfe type algorithm~\cite{Jaggi13revisiting} to obtain both lower and upper bounds on $\tilde{f}(\theta)$. Optimization over $\theta$ is cumbersome, though. We emphasize that it suffices to derive an upper bound of $\f$ in order to derive a valid QEF. For this, we exploit the following result:
\begin{lemma}[Lemma 9 in~\cite{knill2018quantum}]\label{QEFlemma2}
    Suppose given two points in the parameter space, $\theta=(\theta_1,\theta_2),\theta'=(\theta_1+\phi,\theta_2)$ 
    where $\phi\in(0,\pi/2]$, for a point $\theta"=(\theta_1+\varphi,\theta_2)$ such that $\varphi\in[0,\phi]$,
    we have
    \begin{equation}
        \tilde{f}(\theta")\leq \frac{[\sin(\phi-\varphi)+\sin(\varphi)]^{\alpha-1}[\sin(\phi-\varphi)\tilde{f}(\theta)+\sin(\varphi)\tilde{f}(\theta')]}{\sin(\phi)^{\alpha}}\leq\left(\frac{\phi}{\sin(\phi)}\right)^{\alpha}\max(\tilde{f}(\theta),\tilde{f}(\theta')).
    \end{equation}
    A similar result holds when varying the second parameter in the tuple $\theta=(\theta_1,\theta_2)$.
\end{lemma}
To apply this lemma for an upper bound on $\f$, we first divide the parameter space of $\theta$ along the two directions and calculate the values $\tilde{f}(\theta)$ on the mesh grid. It suffices to know an upper bound on $\tilde{f}(\theta')$ where $\theta'=(\theta'_1,\theta'_2)$, $\theta'_1\in [\theta_1, \theta_1+\phi_1]$ and $\theta'_2\in [\theta_2, \theta_2+\phi_2]$. For this, we abbreviate the values
$\tilde{f}(\theta')$ when $\theta'=(\theta_1,\theta_2)$, $(\theta_1+\phi_1,\theta_2)$, $(\theta_1,\theta_2+\phi_2)$ and $(\theta_1+\phi_1,\theta_2+\phi_2)$ by $\tilde{f}_{11}$, $\tilde{f}_{21}$, $\tilde{f}_{12}$ and $\tilde{f}_{22}$ respectively.  According to Lemma~\ref{QEFlemma2}, for any $\theta=(\theta_1+\varphi_1,\theta_2),\,0\leq\varphi_1\leq\phi_1$
we have
\begin{equation}
  \tilde{f}(\theta)\leq\left(\frac{\phi_1}{\sin(\phi_1)}\right)^{\alpha}\max(\tilde{f}_{11}, \tilde{f}_{21}).
\end{equation}
Similarly, for any $\theta=(\theta_1+\varphi_1,\theta_2+\phi_2),\,0\leq\varphi_1\leq\phi_1$
we have
\begin{equation}
  \tilde{f}(\theta)\leq\left(\frac{\phi_1}{\sin(\phi_1)}\right)^{\alpha}\max(\tilde{f}_{12}, \tilde{f}_{22}).
\end{equation}
Then by applying the lemma again along the other direction in the parameter space and in view of the
above two equations, we can determine that
\begin{equation}\label{eq:grid_fmax_upper}
  \tilde{f}(\theta')\leq\left(\frac{\phi_1}{\sin(\phi_1)}\right)^{\alpha}\left(\frac{\phi_2}{\sin(\phi_2)}\right)^{\alpha}\max(\tilde{f}_{11},\tilde{f}_{21},\tilde{f}_{12},\tilde{f}_{22})
\end{equation}
holds for any tuple $\theta'=(\theta'_1,\theta'_2)\in[\theta_1,\theta_1+\phi_1]\times[\theta_2,\theta_2+\phi_2]$. 
At the same time we have
\begin{equation}\label{eq:grid_fmax_lower}
\max(\tilde{f}_{11},\tilde{f}_{21},\tilde{f}_{12},\tilde{f}_{22}) \leq  \tilde{f}(\theta')
\end{equation}
for any tuple $\theta'=(\theta'_1,\theta'_2)\in[\theta_1,\theta_1+\phi_1]\times[\theta_2,\theta_2+\phi_2]$. Therefore, for all $\theta'$ in the region $[\theta_1,\theta_1+\phi_1]\times[\theta_2,\theta_2+\phi_2]$ we can obtain both an upper and a lower bounds on $\tilde{f}(\theta')$. As a consequence, we can obtain both an upper and a lower bound on  $\f$ given a mesh grid. By refining the mesh grid, we can reduce the gap between the lower and upper bounds on $\f$.

In Eq.~\eqref{fmax} we have fixed $\mu(Z)$. In deriving $f_{\text{max}}$ we further take into account the possible bias in $\mu(Z)$ such that $F'(CZ)/f_{\text{max}}$ is a valid QEF for our model. In our setting, we consider that $|\tilde{q}-q|\leq\varepsilon_b$ in the probability distribution of $T_i$, where $\tilde{q}$ is the actual value in the experiment and $q$ is the ideal value. Denote $q_u=q+\varepsilon_b,\,q_l=q-\varepsilon_b$, the set of probability distributions of $Z$ is the convex combination of the probability distributions $\mu_1(Z) = (1-3q_u/4,q_u/4,q_u/4,q_u/4),\,\mu_2(Z) = (1-3q_l/4,q_l/4,q_l/4,q_l/4)$ for $z\in\{00,01,10,11\}$. For the model we consider, any state belonging to it can be expressed as $\tau=\sum_{i=1}^2\lambda_i\tau_i,\,\lambda_i\geq 0,\,\lambda_1+\lambda_2=1$, where $\tau_i\in\mathcal{M}_i(C,Z)=\mu_i(Z)\times\mathcal{M}(C|Z),\,i=1,2$. If $F(CZ)$ is a valid QEF for $\mathcal{M}_i(C,Z),\,i=1,2$, then it is also a valid QEF for $\mathcal{M}(C,Z)$, since for all $\tau=\sum_{i=1}^2\lambda_i\tau_i\in\mathcal{M}(C,Z)$, we have
\begin{equation}
    \sum_{cz}F(cz)\Renyi(cz|z)_{\tau_E(cz)}\leq\sum_{cz}F(cz)\sum_{i=1}^2\Renyi(cz|z)_{\tau_{i,E}(cz)}\leq\lambda_1+\lambda_2=1.
\end{equation}
With this conclusion, we can calculate the regularising factor $\f^{i}$ for $\mathcal{M}_i(C,Z),\,i=1,2$ such that $F'(CZ)/\f^{i}$ is a valid QEF for both models $\mathcal{M}_i(C,Z)$. By taking $\f=\max\{\f^{1},\,\f{^2}\}$ we can derive a valid QEF for $\mathcal{M}(C,Z)$, that is, $F(CZ)=F'(CZ)/\f$.

\subsection{Randomness extraction}\label{extraction}
We use Toeplitz extractor in the experiment, which takes the experimental output as input and delivers a sequence of near uniform random bits~\cite{Impagliazzo:Leftover:1989, frauchiger2013true, Ma13, liu2018device}. The Toeplitz extractor is a quantum-proof strong extractor, defined as follows:

\begin{definition}[Quantum-Proof Strong Extractor~\cite{nisan1996randomness,konig2011sampling,de2012trevisan,Ma13}] A function $\text{Ext: }\{0,1\}^n\times\{0,1\}^d\rightarrow\{0,1\}^m$ is a quantum-proof $(k,\varepsilon_x)$-strong extractor with a uniform seed, if for all classical-quantum states $\rho_{XE}$ classical on $X$ with $H_{\text{min}}(X|E)_\rho\geq k$ and a uniform seed $Y$, we have
\begin{equation}
    \dfrac{1}{2}\|\rho_{\text{Ext}(X,Y)YE}-\rho_{U_m}\otimes\rho_Y\otimes\rho_E\|_1\leq\varepsilon_x,
\end{equation}
where $\|\cdot\|_1$ is the trace norm, $\rho_{U_m}=\frac{I}{2^m}$ is the fully mixed state of dimension $2^m$ and $\rho_Y$ is the fully mixed state of the seed $Y$.
\end{definition}

The definition states that, given a quantum-proof $(k,\varepsilon_x)$-strong extractor with a uniform seed, $m$ bits of uniformly distributed random bits can be provided (except for a failure probability no larger than $\varepsilon_x$) if there is a guarantee of $k$ bits of min-entropy in the input $X$.
For brevity, we call $\varepsilon_x$ the security parameter of the extractor. This definition naturally guarantees a composability property, shown by the following lemma:
\begin{lemma}[\cite{de2012trevisan,renner2008security}]
  If $\text{Ext: }\{0,1\}^n\times\{0,1\}^d\rightarrow\{0,1\}^m$ is a quantum-proof $(k,\varepsilon_x)$-strong extractor, then for any classical-quantum state $\rho_{XE}$ such that $H_{\text{min}}^{\varepsilon_h}(X|E)_\rho\geq k$, we have
\begin{equation}\label{Eq:Composable}
    \dfrac{1}{2}\|\rho_{\text{Ext}(X,Y)YE}-\rho_{U_m}\otimes\rho_Y\otimes\rho_E\|_1\leq\varepsilon_x+2\varepsilon_h.
\end{equation}
\end{lemma}

From this property, we can reuse the extraction seed albeit a linear increase in the security parameter.

An $m\times n$ Toeplitz matrix takes the from,
\begin{equation}\label{}
  T_{m\times n} =
  \left(
    \begin{array}{ccccc}
    a_0     & a_{-1}  & \cdots & a_{-(n-2)}   & a_{-(n-1)} \\
    a_1     & a_0     & \ddots &              & a_{-(n-1)+1} \\
    a_2     & a_1     & \ddots & \ddots       & \vdots     \\
    \vdots  & \vdots  &        & \ddots       & a_{-(n-1)+(m-2)} \\
    a_{m-1} & a_{m-2} & \cdots & a_{-n+(m-1)} & a_{-(n-1)+(m-1)} \\
    \end{array}
  \right).
\end{equation}
For the Toeplitz matrix, we have the following lemma that guarantees its use as a quantum-proof strong extractor:
\begin{lemma}[\cite{renner2008security}]
  The set of all Toeplitz matrices can be used as a quantum-proof strong $(k,\varepsilon_x)$ extractor with $\varepsilon_x=2^{-(k-m)/2}$, where $m$ is the output length.
\end{lemma}

The experimental output is written in the form of an $n-$dimensional vector,
\begin{equation}\label{}
  V_{n} =
  \left(
    \begin{array}{c}
    v_0 ,
    v_1 ,
    v_2 ,
    \cdots  ,
    v_{n-1} \\
    \end{array}
  \right)^T.
\end{equation}
The output of the Toeplitz extractor is a sequence of nearly uniform random bits $U_m$, with $R_{m}=T_{m\times n}\times V_{n}$, which is given as,
\begin{equation}\label{}
  U_{m} =
  \left(
    \begin{array}{c}
    u_0 ,
    u_1 ,
    u_2 ,
    \cdots  ,
    u_{m-1} \\
    \end{array}
  \right)^T.
\end{equation}

\subsection{Protocol Security}\label{Supp:SecDef}
In this subsection, we shall discuss the security of the protocol under the framework of quantum probability estimation. In this work we use a composable security definition~\cite{canetti2000security,ben2004general,portmann2014cryptographic}:

\begin{definition}[Protocol soundness and completeness]
  A random number generation protocol with an $m$-bit output $\bm{C}$ is called $(\varepsilon_S,\varepsilon_C)$-secure if it is\\
  (1) $\varepsilon_S$-Sound: The output satisfies
  \begin{equation}
    \dfrac{1}{2}p_{\Phi}\|\rho_{\bm{C}E|\Phi} - \tau_m\otimes\rho_{E|\Phi}\|_1\leq\varepsilon_S,
  \end{equation}
  where $\Phi$ represents the event that the protocol does not abort, $\rho_{\bm{C}E|\Phi}$ is the normalised final state conditioned on a success, with $\rho_{\bm{C}}$ giving the output $\bm{C}$, and $\rho_E$ the system of side information, and $\tau_m$ is a maximally mixed state of $m$ qubits. The probability that the protocol does not abort is $p_{\Phi}$. \\
  (2) $\varepsilon_C$-Complete: There exists an honest implementation where the protocol
  observes an identical and independent behaviour, and the probability that it does not abort satisfies $p_{\Phi}\geq 1-\varepsilon_C$.
\end{definition}

In our experiment, we apply the following protocol for random number generation~\cite{knill2018quantum}:

\begin{tcolorbox}[title = {Box 1. Input-Conditional Random Number Generation}]
\textbf{Input: }

$k_{\text{gen}}$: the number of random bits to be generated. \\
$\varepsilon_S$: the soundness error.\\

\textbf{Given: }\\
An experiment of $N$ trials, with input $\bm{Z}$ and output $\bm{C}$ of length $2N$.\\
A valid QEF $F(\bm{CZ})$ with power $\alpha$ for the model of the experiment $\mathcal{M}(\bm{C},\bm{Z})$.\\
A quantum-proof $(k,\varepsilon_x)$-strong extractor $\text{Ext: }\{0,1\}^{2N}\times\{0,1\}^d\rightarrow\{0,1\}^{k_{\text{gen}}}$.\\

\textbf{Output: }\\
A bit string of length $k_{\text{gen}}$.

\tcblower
\textbf{Procedures:}
\begin{enumerate}
\item Assign the smoothing parameter in randomness generation $\varepsilon_h$, the security parameter of the extractor
    $\varepsilon_x$ such that $\varepsilon_S = 2\varepsilon_h + \varepsilon_x$.
\item Assign the success threshold of the experiment $h_s$.
\item Obtain a realisation $s$ of the uniform seed $S$ of length $d$.
\item Perform the randomness generation experiment and obtain a realisation $\bm{c},\bm{z}$ of $\bm{C},\bm{Z}$ and the QEF value $F(\bm{cz})$.
\item If $F(\bm{cz})\geq2^{h_s}(\alpha-1)$, the protocol succeeds and return $U_{k_{\text{gen}}} = \text{Ext}(\bm{c},s)$; otherwise, the protocol fails and return $U_{k_{\text{gen}}} = 0^{\frown k_{\text{gen}}}$.
\end{enumerate}

\end{tcolorbox}

Under the QPE framework, the soundness and completeness for the protocol has been proved (see Sec. IIIC of~\cite{knill2018quantum}).
We mention that in this work, the total soundness error in the soundness proof in~\cite{knill2018quantum} shall be modified to $\varepsilon_S = \varepsilon_x + 2\varepsilon_h$ as given in Eq.~\eqref{Eq:Composable}, since the soundness definition used in this work is slightly different from that in~\cite{knill2018quantum}.

\section{Parameter Determination}\label{sec:ParDet}
Before the execution of randomness expansion, we need to determine a probability distribution of inputs that supports the task, set the largest allowed number of experimental trials, the soundness error and the failure probability of the protocol, and find a valid QEF that yields a good randomness expansion rate. In our experiment, we determine these parameters with the following procedure:
\begin{tcolorbox}[title = {Box 2. Procedure for parameter determination:}]
	1. Determine an input probability distribution $\mu(Z)$ that supports randomness expansion.\\

	2. Set the input probability distribution as $\mu(Z)$. Conduct a series of ``training trials'' consuming $k_0$ bits of entropy and determine an empirical input-output probability distribution $\nu(CZ)$ for randomness expansion.\\
	
	3. Determine a QEF $F(CZ)$ with power $\alpha$ that yields a large $r_\nu(F,\alpha)$.\\

    4. Set the randomness expansion target $k_{\text{exp}}$, the largest allowed number of trials $N$, the soundness error in randomness expansion $\varepsilon_h$, the soundness error in randomness extraction $\varepsilon_x$, the completeness error of the protocol $\varepsilon_C$, and a lower bound on the success probability $\kappa$ of the protocol.\\

    5. Set the success threshold in the randomness expansion experiment $h_s$.
\end{tcolorbox}

In Step 1, based on our experimental parameters, we simulate an input-conditional probability distribution $\bar{\nu}(C|Z)$ using the Eberhard model~\cite{Eberhard93}, and construct the model for the joint probability distribution $\bar{\nu}(CZ)$ of the input and output by multiplying different input distributions $\mu(Z)$ under the spot-checking protocol. We consider a simulated input setting to be feasible for randomness expansion if $r_{\bar{\nu}}(F,\alpha)>r_{\text{in}}$ for some QEF $F(CZ)$ with power $\alpha$.
In this step no randomness shall be consumed. We take the assumption that the simulated input-output probability distributions do not deviate from the actual experimental behaviour too much.

To determine the input probability distribution and QEF for an efficient randomness expansion, for the best we should minimize the expected number of trials for randomness expansion.
Nonetheless, a suboptimal solution can be accepted. In our determination, under a given simulated probability distribution $\bar{\nu}(CZ)$,
we vary the value $\alpha$, and for each fixed value of $\alpha$,
in principle we can optimize the QEF that yields a largest expected randomness rate witnessed by QEF [Eq.~\eqref{QEFopt}].
As stated, however, the optimization is difficult to tackle. In practice, we make use of the PEF, and consider the optimization in Eq.~\eqref{PEFoptTsirelson}.
For the same input-output probability distribution $\bar{\nu}(CZ)$, the expected output randomness rate witnessed by the PEF $F'(CZ)$ is $r_{\bar{\nu}}(F',\alpha)=\mathbb{E}_{\bar{\nu}}\log_2(F'(CZ))/(\alpha-1)$. We assume $r_{\bar{\nu}}(F',\alpha)$ is close to $r_{\bar{\nu}}(F,\alpha)$, and we will show that it is reasonable in our experiment.
In our experiment, we determine the input probability distribution in reference to the value of $r_{\bar{\nu}}(F',\alpha)$.

With the input probability distribution fixed, following the route suggested in~\cite{Zhang2018Low}, we carry out a series of training trials under this input setting and obtain an empirical probability distribution $\nu(CZ)$ adapted to the model $\mathcal{M}(C,Z)$ we use.
Still, we first search for a PEF $F'(CZ)$ with power $\alpha$ that might support the randomness expansion efficiently. Then we scale the PEF $F'(CZ)$ with a parameter $f_{\text{max}}$, such that $F'(CZ)/f_{\text{max}}$ becomes a valid QEF for the model against quantum side information with the same power, following the approach given in Sec.~\ref{supp:opt_qef_sec}.


Afterwards, we determine
the largest allowed number of experimental trials $N$ with the aid of an honest protocol that observes an i.i.d.behaviour.
We note that the assumption is only used in parameter determination. In the actual randomness expansion experiment we do not make any assumption about the input-output distribution.
In the honest protocol with the input-output probability distribution of $\nu(CZ)$,
we set the smoothing parameter in randomness generation to be $\varepsilon_h$,
and the preset success probability of the protocol to be $\gamma$.
Here $\gamma = 1 - \varepsilon_C$, where $\varepsilon_C$ is the completeness error of the protocol.
By the central limit theorem, the distribution of the variable $N\log_2(F(CZ))/(\alpha-1)$ can be well approximated by a normal distribution, with the mean $Nr_{\nu}(F,\alpha)$ and the variance $N\sigma_\nu^2$, where $\sigma_\nu$ is the standard deviation of the random variable $\log_2(F(CZ))/(\alpha-1)$ with respect to the input-output distribution $\nu(CZ)$, determined by
\begin{equation}
  \sigma_\nu^2 = \mathbb{E}_\nu\left[\left(\dfrac{\log_2(F(CZ))}{\alpha-1}\right)^2\right] - \mathbb{E}_\nu\left[\left(\dfrac{\log_2(F(CZ))}{\alpha-1}\right)\right]^2.
\end{equation}
The success probability of the honest protocol is given by
\begin{equation}\label{Eq:SuccProb}
 \gamma = Q(-[Nr_{\nu}(F,\alpha)-\bar{h}_s]/(\sqrt{N}\sigma_\nu)),
\end{equation}
where $Q$ is the tail distribution function, and $\bar{h}_s$ is the success threshold for an honest protocol. Here we use a barred notation to distinguish the protocol success probability from the one for the adversarial condition $\kappa$ that is used in the actual experiment.
According to Eq.~\eqref{eq:QEFRandomness}, with an honest protocol, to generate at least $k$ bits of randomness in $N$ trials in the randomness expansion experiment (before the randomness extraction), the success threshold is determined by
\begin{equation}\label{Eq:HonestThreshold}
  \bar{h}_s= k + \frac{1}{\alpha-1}\log_2\left(\frac{2}{\varepsilon_h^2}\right) + \frac{\alpha}{\alpha-1}\log_2\left(\frac{1}{\gamma}\right).
\end{equation}
For the expansion task to expand at least $k_{\text{exp}}$ bits of near-uniform random bits at the end of the protocol, $k$ is determined by
\begin{equation}\label{Eq:ExpansionThreshold}
  k = k_0 + k_{\text{exp}} + Nr_{\text{in}} - 2\log_2\varepsilon_x = k_{\text{gen}} - 2\log_2\varepsilon_x.
\end{equation}
$Nr_{\text{in}}$ the largest amount of randomness that might be consumed in input setting, and the term $-2\log_2\varepsilon_x$ is required by the randomness extraction procedure.
After the randomness extraction there are $k_{\text{gen}}$ near-uniform random bits generated (without the deduction of the entropy cost, which is $k_0 + Nr_{\text{in}}$ in total).
After setting the completeness error $\varepsilon_C$ and the determination of the protocol success probability for an honest protocol $\gamma$, the soundness error in randomness expansion $\varepsilon_h$,
we determine the largest allowed number of trials in randomness expansion $N$ by solving Eq.~\eqref{Eq:SuccProb}\eqref{Eq:HonestThreshold} and~\eqref{Eq:ExpansionThreshold} jointly.

To determine the success threshold $h_s$ for the actual randomness expansion experiment, we need to set the lower bound to the actual protocol success probability $\kappa$. We emphasize that now we do not assume the experimental behaviour to be honest.
Literature suggests taking $\kappa=\varepsilon_h$ to obtain a conservative lower bound on the $\varepsilon_h$-smooth min-entropy~\cite{knill2018quantum}.
Then the success threshold is determined by
\begin{equation}\label{Eq:ActualThreshold}
  {h}_s= k + \frac{1}{\alpha-1}\log_2\left(\frac{2}{\varepsilon_h^2}\right) + \frac{\alpha}{\alpha-1}\log_2\left(\frac{1}{\kappa}\right).
\end{equation}

\section{System characterization}\label{Supp:Experiment}
\begin{figure*}[hbt!]
	\centering
	\resizebox{15cm}{!}{\includegraphics{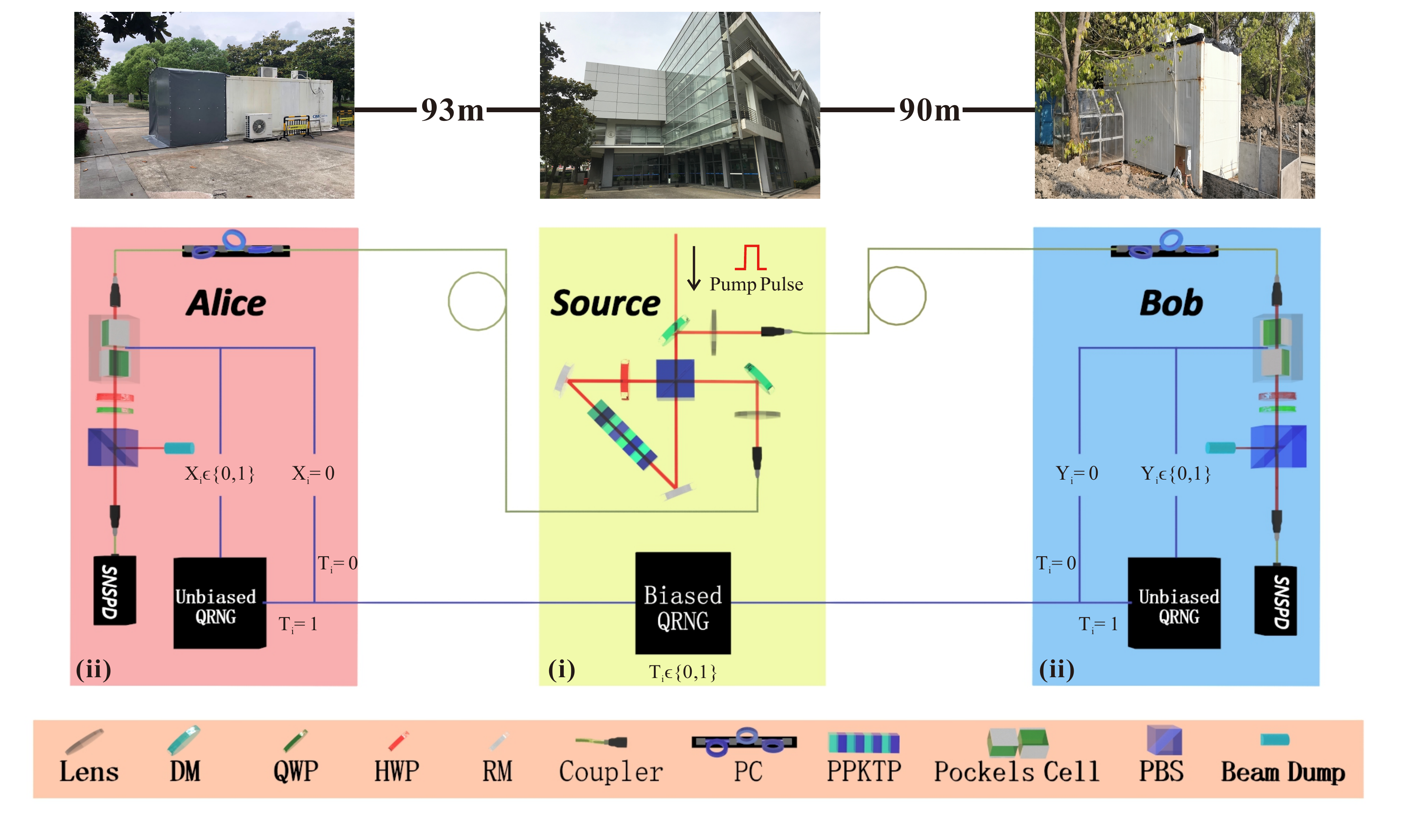}}
	\caption{
        Schematic illustration of our experimental realization with entangled photons (see \cite{liu2018device} for details). (i) Creation of pairs of entangled photons: Light pulses of 10 ns are injected into a periodically poled potassium titanyl phosphate (PPKTP) crystal in a Sagnac loop to generate polarization-entangled photon pairs at 1560 nm, which are sent to measurement stations Alice and Bob.
        (ii) Realization of single photon polarization state measurement with a group of polarization optics: Polarization controller (PC), Pockels cell, quarter-wave plate (QWP), half-wave plate (HWP) and polarizing beam splitter (PBS), and superconducting nanowire single-photon detectors (SNSPDs). A biased quantum random number generator (QRNG) generates random bit $T_i\in\{0,1\}$. Alice and Bob each has a local unbiased QRNG to set the measurement base settings. RM and DM are reflecting and dichorism mirrors. The laboratories for the experimental set-ups of Alice, Bob and the entanglement source are separated in space (see Appendix~\ref{spacetime} for details).
	}
	\label{Fig:Experiment}
\end{figure*}

In Fig.~\ref{Fig:Experiment} we present a schematic illustration of our experimental set-up. The set-up is similar to the one in our previous experimental work of DIQRNG in~\cite{liu2018device}, and the technical details can be found therein.
In this section we discuss the determination of the essential experimental parameters of this work.

\subsection{Assignment of input probability distribution}\label{app:assign}

\begin{figure}[!hbt]
    \centering
    \resizebox{10cm}{!}{\includegraphics{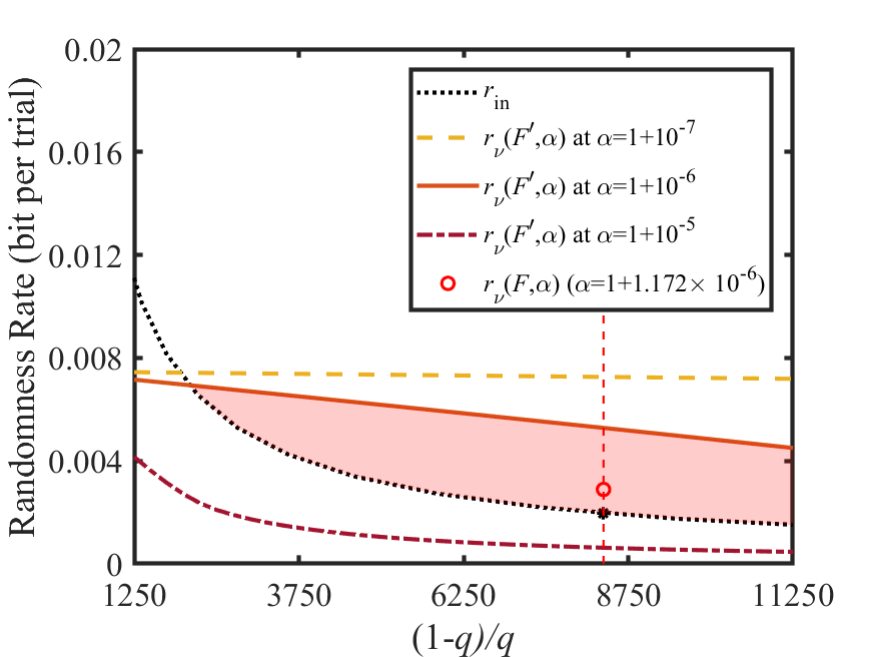}}
    \caption{The simulated expected randomness rate in the spot-checking protocol (PEF). To determine the input setting for randomness expansion, we simulated a set of joint probability distributions with different biased input probability distributions, and optimize the expected output randomness rate determined by PEFs. The black dotted line gives the input entropy rate with repect to different $b_s$, and the other three lines (yellow dash, orange solid, maroon dot-dash lines) give the expected output randomness rate determined by optimized PEFs with different powers. The estimated feasible region for randomness expansion is determined by $r_\nu(F',\alpha)>r_{\text{in}}$, and we mark the area in pink for the power $\alpha=1+10^{-6}$ as an example.
    From the simulation result and the test behaviour of the biased random number generator, we determine a setting $(1-q)/q=8375$ (shown by the red dotted vertical line) for randomness expansion. We also mark the value $r_\nu(F,\alpha)$ witnessed by the QEF we utilise in the experiment (red open circle) for comparison, which is determined from the training trials.
    }
    \label{Fig:spot-check PEF}
\end{figure}

In this subsection we present the results in determining the input-output probability distribution.
Following the route in Sec.~\ref{sec:ParDet},
we simulate a set of input-output distributions $\bar{\nu}(CZ)=\mu(Z)\bar{\nu}(C|Z)$ with different input probability distributions $\mu(Z)$, and $\bar{\nu}(C|Z)$ is an input-conditioned probability distribution simulated based on our previous experimental data.
In the simulation of a set of input-conditioned probability distribution with the Eberhard model, the key experimental parameters for the simulation include the efficiencies in different processes of the experiment, the choices of measurement bases and the entangled state, the state fidelity, the mean photon number of the entangled photon source, the visibility, etc.
For the detailed simulation procedure, one may refer to the Supplementary Material of~\cite{LiPRL2018}.
As shown by Fig.~\ref{Fig:spot-check PEF}, we optimize the PEF to maximize $r_{\bar{\nu}}(F',\alpha)$, with various values of $(1-q)/q$, which is the ratio of the number of spot trials to the number of checking trials, and the order $\alpha$ in the QEF. If $r_{\bar{\nu}}(F',\alpha)>r_{\text{in}}$ we consider the input probability distribution to be feasible for randomness expansion. We try to find {an input probability distribution} to minimize the number of trials for randomness expansion
to our best.
From PEF results, we find that with the power $\alpha$ around the level of $1+1\times10^{-6}$, an efficient randomness expansion is possible. Afterwards we perform a fine-grained optimization with $\alpha$ varying around this value, and determine $\alpha=1+1.172\times 10^{-6}$.
From the PEF optimization results, we determine the ratio of the number of spot trials to the number of checking trials to be $(1-q)/q=8375 (q=0.000119)$, corresponding to an input entropy rate of $r_{\text{in}}=0.00197$. A detailed description of experimentally realising a biased QRNG can be referred to our previous work~\cite{liu2018device}.


\subsection{Determination of single photon efficiency}\label{efficiency}
We determine the single photon heralding efficiency as $\eta_A=C/N_B$ and $\eta_B=C/N_A$ for Alice and Bob, in which two-photon coincidence events $C$ and single photon detection events for Alice $N_A$ and Bob $N_B$ are measured in the experiment, which are listed in Tab.~\ref{tab:OptEffAB}.

In this table, $\eta^{sc}$ is the efficiency to couple entangled photons into single mode optical fibre,  $\eta^{so}$ is the efficiency for photons passing through the optical elements in the source, $\eta^{fibre}$ is the transmittance of fibre connecting source to measurement station, $\eta^{m}$ is the efficiency for light passing through the measurement station, and $\eta^{det}$ is the single photon detector efficiency. $\eta^{so}$, $\eta^{fibre}$, $\eta^{m}$, $\eta^{det}$ can be measured with classical light beams and NIST-traceable power meters.
\begin{table}[htb]
\centering
  \caption{Optical efficiencies in the experiment.}
\begin{tabular}{ccccccc}
\hline
 Parties\, & Heralding,\,$\eta$\, & $\eta^{sc}$\, & $\eta^{so}$\, & $\eta^{fibre}$\, & $\eta^{m}$\, & $\eta^{det}$ \\
\hline
Alice  & 80.50\% & 92.3\% & \multirow{2}{*}{95.9\%} & \multirow{2}{*}{99\%} & 95.1\% & 96.6\%\\
Bob    & 82.20\%& 93.1\% &                       &                       & 95.3\% & 97.3\% \\
\hline
\end{tabular}
\label{tab:OptEffAB}
\end{table}

\subsection{Quantum state and measurement bases}\label{sec:measure}
To maximally violate the Bell inequality in experiment, we create non-maximally entangled two-photon state~\cite{Eberhard93} $\cos(24.56^\circ)\ket{HV}+\sin(24.56^\circ)\ket{VH}$ and set measurement bases to be  $A_1=-83.02^\circ$ and $A_2=-118.58^\circ$ for Alice, and $B_1=6.98^\circ$ and $B_2=-28.58^\circ$ for Bob, respectively.

We measure diagonal/anti-diagonal visibility in the bases set ($45^\circ, -24.56^\circ$), ($114.56^\circ, 45^\circ$) for minimum coincidence, and in the bases set ($45^\circ, 65.44^\circ$), ($24.56^\circ, 45^\circ$) for maximum coincidence, where the angles represent measurement basis  $\cos(\theta)\ket{H}+\sin(\theta)\ket{V}$ for Alice and Bob. By setting the mean photon number to $\mu=0.0025$ to suppress the multi-photon effect, we measure the visibility to be $99.5\%$ and $98.4\%$ in horizontal/vertical basis and diagonal/anti-diagonal basis.

\begin{figure}[htb]
\centering
    \subfigure[]{
      \includegraphics[width=8cm]{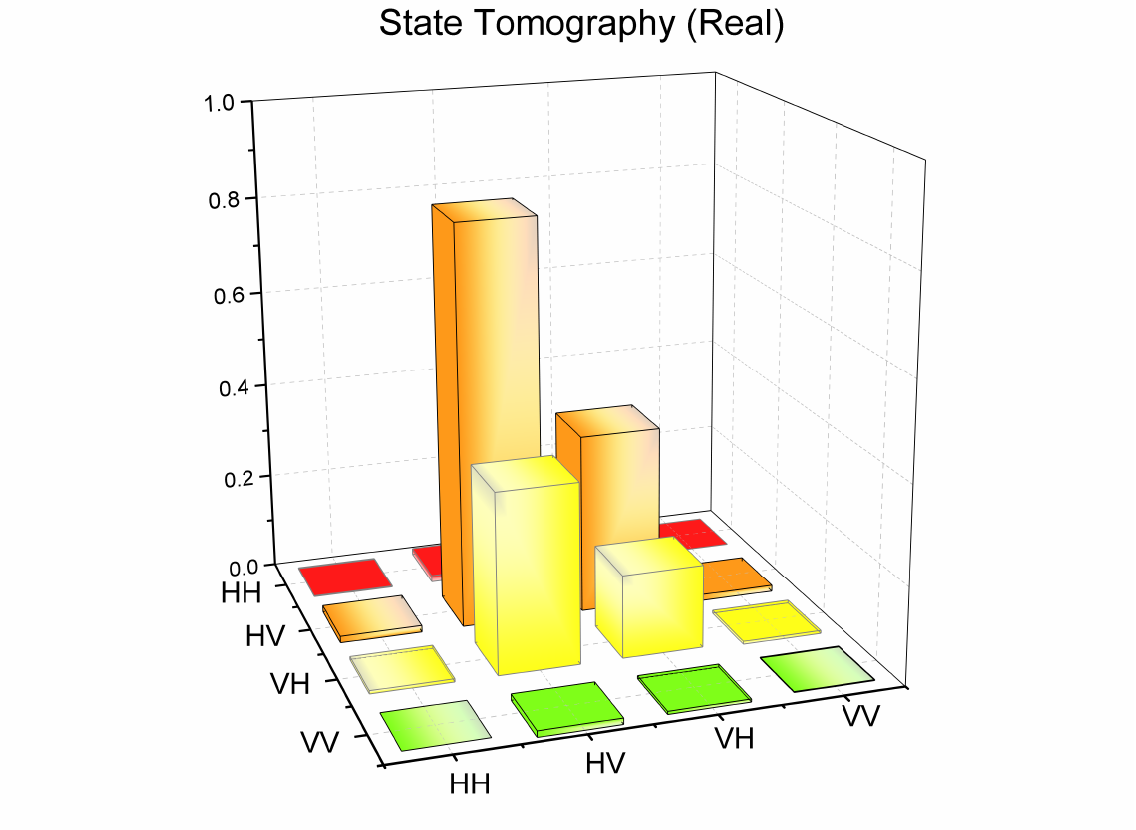}
    }
    \subfigure[]{
      \includegraphics[width=8cm]{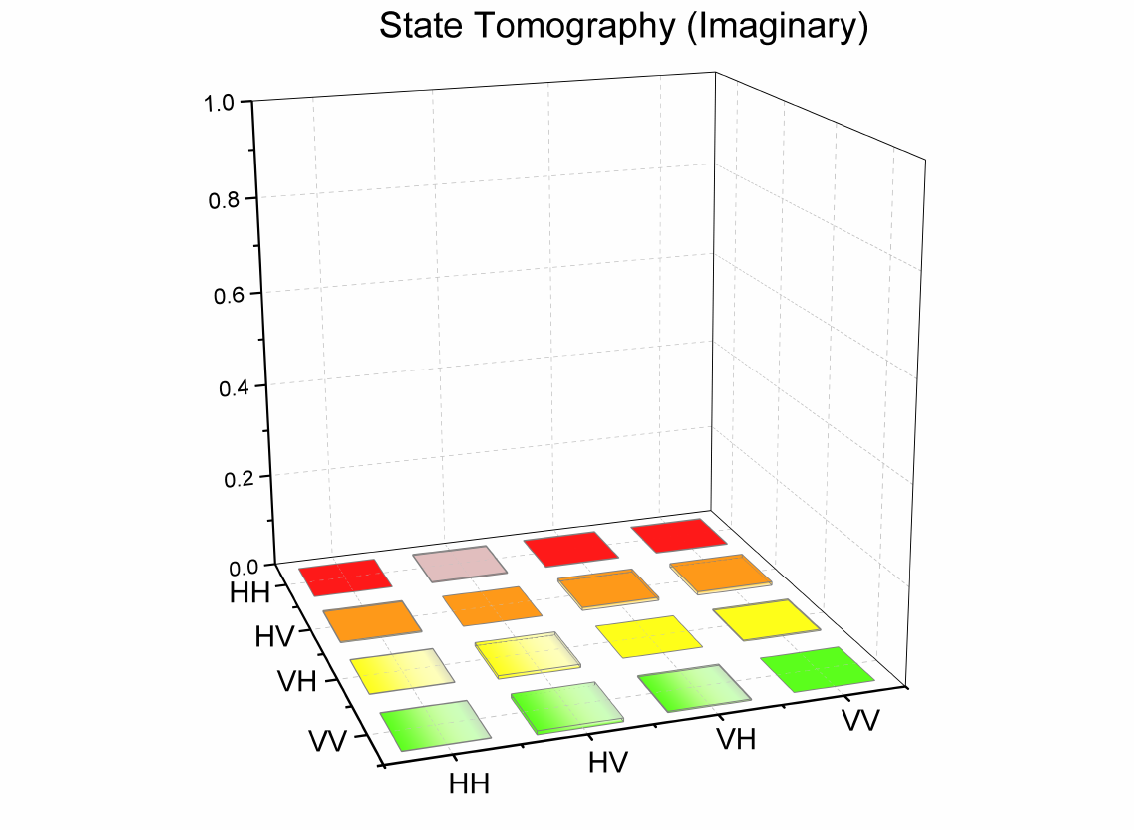}
    }
\caption{(color online)
Tomography of the produced two-photon state in the experiment, with real and imaginary components shown in (a) and (b), respectively.}
\label{Fig.Tomo}
\end{figure}

We perform quantum state tomography measurement of the non-maximally entangled state, with result shown in Fig.~\ref{Fig.Tomo}. The state fidelity is $99.16\%$. We attribute the imperfection to multi-photon components, imperfect optical elements, and imperfect spatial/spectral mode matching.

\subsection{Spacetime configuration of the experiment}\label{spacetime}
\begin{figure}[tbh]
    \centering
    \resizebox{10cm}{!}{\includegraphics{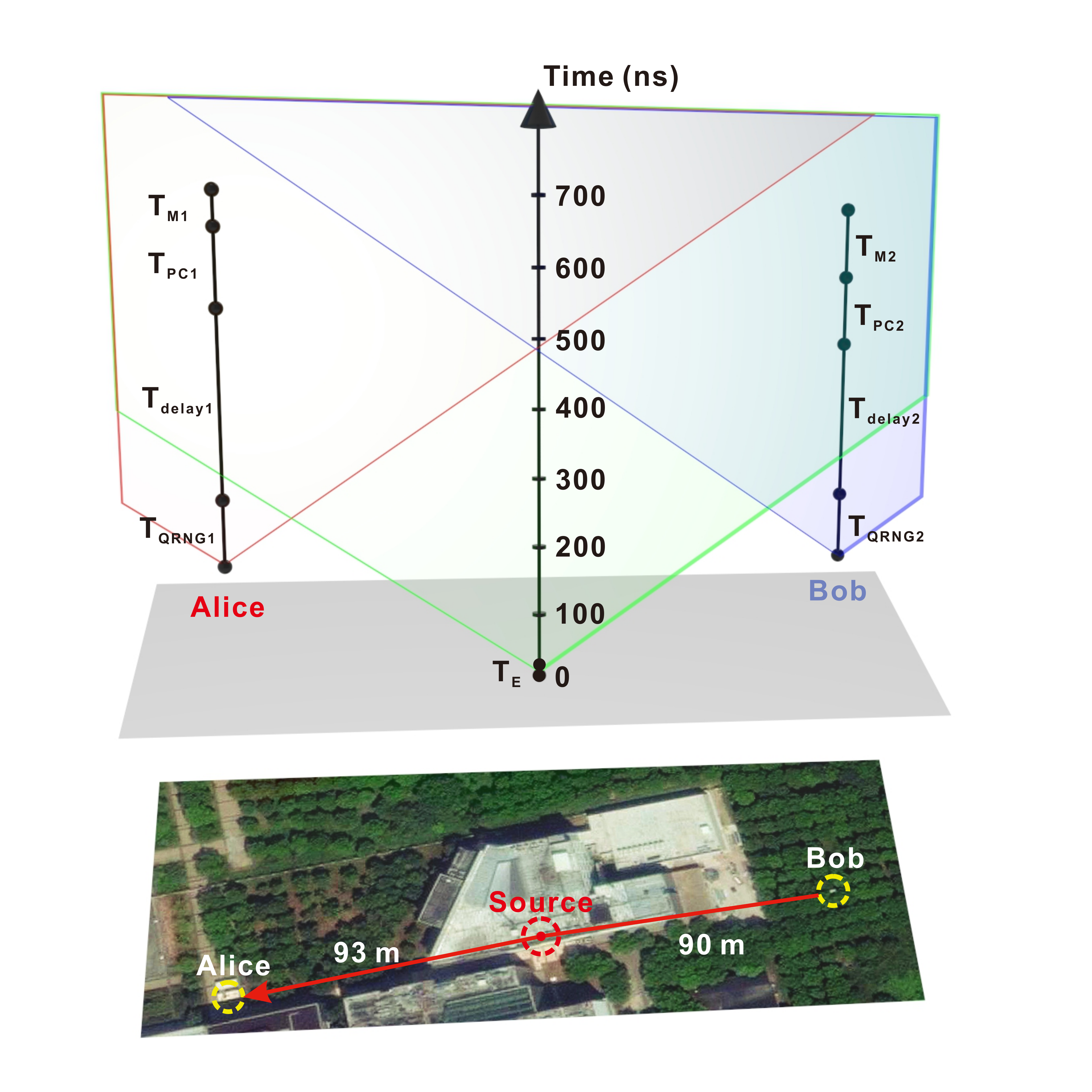}}
    \caption{Schematics of experimental configuration. Bottom: creation of a pair of entangled photons at the source and measurement of photons at stations A and B; upper: corresponding spacetime analysis exhibiting spacelike separation between relevant events, drawn to the scale. The time segments correspond to time elapse for: $T_{E}-$generation of a pair of entangled photons at the source; $T_{QRNG1,2}-$generation of random bits from unbiased quantum random number generator (QRNG) (1,2$-$station A,B); $T_{Delay1,2}-$delay between QRNG and Pockcels cell; $T_{PC1,2}-$Pockcels cell gets ready for state measurement after receiving a random bit; $T_{M1,2}-$photon detector outputs an electronic signal.
    }
    \label{Fig:spacetime}
    \end{figure}

To close the locality loophole, space-like separation must be satisfied between relevant events at Alice and Bob's measurement stations: the state measurement events by Alice and Bob, measurement event at one station and the setting choice event at the other station (Fig.~\ref{Fig:spacetime}). We then obtain

\begin{equation}
	\begin{cases}
(|SA| + |SB|) / c > T_E - (L_{SA} - L_{SB}) / c + T_{QRNG1} + T_{Delay1} + T_{PC1} +T_{M2}, \\
(|SA| + |SB|) / c > T_E + (L_{SA} - L_{SB}) / c + T_{QRNG2} + T_{Delay2} + T_{PC2} +T_{M1},
	\end{cases}
\label{Eq:SC1}
\end{equation}
where $|SA|$ = 93 m ($|SB|$ = 90 m) is the free space distance between entanglement source and Alice's (Bob's) measurement station, $T_E$ = 10 ns is the generation time for entangled photon pairs, which is mainly contributed by the 10 ns pump pulse duration, $L_{SA}$ = 191 m ($L_{SB}$ = 173.5 m) is the effective optical path which is mainly contributed by the long fibre (130 m, 118 m) between source and Alice/Bob's measurement station, $T_{QRNG1}=T_{QRNG2}$ = 96 ns is the time elapse for unbiased QRNG to generate a random bit, $T_{Delay1}$ = 270 ns ($T_{Delay2}$ = 230 ns) is the delay between unbiased QRNG and Pockels cells, $T_{PC1}$ = 112 ns ($T_{PC2}$ = 100 ns) including the internal delay of the Pockcels Cells (62 ns, 50 ns) and the time for Pockcels cell to stabilize before performing single photon polarization state projection after switching which is 50 ns, which implies that the experimental time is able to be shortened by increasing the repetition rate of the experiment because the small $q$ reduces the impact of the modulation rate of the Pockels cells, $T_{M1}$ = 55 ns ($T_{M2}$ = 100 ns) is the time elapse for SNSPD to output an electronic signal, including the delay due to fibre and cable length.

Measurement independence requirement is satisfied by space-like separation between entangled-pair creation event and setting choice events, so we can have

\begin{equation}
	\begin{cases}
|SA| / c > L_{SA} / c  - T_{Delay1} - T_{PC1}\\
|SB| / c > L_{SB} / c  - T_{Delay2} - T_{PC2}
	\end{cases}
\label{Eq:SC2}
\end{equation}

As shown in Fig.~\ref{Fig:spacetime}, Alice's and Bob's random bit generation events for input setting choices are outside the future light cone (green shade) of entanglement creation event at the source.

\section{Experimental Results}\label{app:results}

\subsection{Randomness expansion task: 512 bits}\label{512}

Our DIQRE task is set to outperform the input by at least 512 near-uniform random bits at the end of the protocol.
We set the soundness errors $\varepsilon_h = 2^{-32},\,\varepsilon_x = 2^{-100}$, the success probability for an honest protocol $\gamma = 99.3\%$ (equivalent to the one-sided $2.5\sigma$ criterion and corresponding to the completeness error $\varepsilon_C = 0.7\%$), and the lower bound to the success probability in the actual experiment to be the same as the soundness error in randomness expansion, $\kappa = \varepsilon_h = 2^{-32}$.
We carry out three hours of training trials under the input setting determined in Sec.~\ref{app:assign}, which corresponds to a consumption of $k_0\approx8.50\times10^7$ bits of entropy, and obtain an empirical probability distribution $\nu(CZ)$ (see Table~\ref{table:EmpiricalData} and~\ref{table:EmpiricalDist}).
With respect to $\nu(CZ)$, we optimize the QEF value according to the methods in Sec.~\ref{theory}.
We first optimize a PEF under the no-signaling condition and Tsirelson's bounds. The power $\alpha$ of the optimized PEF is $\alpha = 1 + 1.172\times 10^{-6}$, and the values of the PEF are given in Table~\ref{table:PEF}.
We derive the value of the PEF using the $\text{C++}$ programming language and the package \emph{float128}, attaining a precision to $35$ decimal places. After obtaining the PEF, we normalize it and solve the optimization of $\f$ in Eq.~\eqref{fmax}. Such an optimization problem is tackled via the parallel computation toolbox in Matlab. The overall QEF rescaling factor is the multiplication of the sum of the 16 PEF values and $\f$. We derived an upper bound of $1+1.12\times 10^{-9}$ to the overall rescaling factor.
The QEF value can be obtained by dividing the PEF with this factor.
The expected output entropy rate witnessed by the QEF is $r_{\nu}(F,\alpha)=0.00289$. For the optimization of $\f$, we used 24 workers for parallel computation, taking $30963.214420$ seconds.
With these parameters, we determine the largest allowed number of trials in randomness expansion to be $N=2.35\times 10^{11}$, and the success threshold in the experiment to be $h_s = 6.31\times10^8$ bits.

%

\begin{table}[htb]
\centering
\caption{The empirical input-output counts from the training set. 
}
\label{table:EmpiricalData}
\begin{tabular}{c|cccc}
\hline
\diagbox[width=5em,trim=l]{(x,y)}{(a,b)} & 00 & 01 & 10 & 11\\
\hline
00 & 42212881971 & 318991793 & 275003068 & 629231576 \\
01 & 1240932 & 27956 & 6070 & 21021 \\
10 & 1243249 & 6689 & 27566 & 21427 \\
11 & 1201874 & 45577 & 45611 & 3620 \\
\hline
\end{tabular}
\end{table}

\begin{table}[htb]
\centering
\caption{The empirical input-output distribution $\nu(ABXY)$. An MLE is applied to the empirical data to derive a probability distribution adapted to the model used. Here we present the result to 20 decimal places.
}
\label{table:EmpiricalDist}
\begin{tabular}{c|cccc}
\hline
\diagbox[width=5em,trim=l]{(x,y)}{(a,b)} & 00 & 01 & 10 & 11\\
\hline
00 & 0.97199465625472059038 & 0.00715304637435085818 & 0.00631932133640711064 & 0.01444343448724344329 \\
01 & 0.00002857430319592834 & 0.00000065311397457615 & 0.00000014282560923430 & 0.00000047693964624017 \\
10 & 0.00002858371146706307 & 0.00000015598344257690 & 0.00000061881913175867 & 0.00000048866838458032 \\
11 & 0.00002769260884464488 & 0.00000104708606499509 & 0.00000102451996051776 & 0.00000008296755582123 \\
\hline
\end{tabular}
\end{table}

\begin{table}[htb]
\centering
\caption{The optimal PEF $F'(ABXY)$ with power $\alpha=1+1.172\times10^{-6}$. Here we present the result to 20 decimal places.}
\label{table:PEF}
\begin{tabular}{c|cccc}
\hline
\diagbox[width=5em,trim=l]{(x,y)}{(a,b)} & 00 & 01 & 10 & 11\\
\hline
00 & 1.00000000110510334216 & 0.99999903566359593654 & 0.99999908430264417003 & 1.00000100579637485331 \\
01 & 1.00022934253952033856 & 0.98995503866430756279 & 0.93407594075304811731 & 1.02073956349038930113 \\
10 & 1.00023612022915342478 & 0.93601590290081493339 & 0.98948855714127381677 & 1.02175557026355190437 \\
11 & 0.99949697803240911131 & 1.00975800133726889562 & 1.01028078627036155268 & 0.92372918497532785497 \\
\hline
\end{tabular}
\end{table}


\subsection{System Robustness}\label{system}
\begin{figure}[tbh]
    \centering
    \resizebox{16cm}{!}{\includegraphics{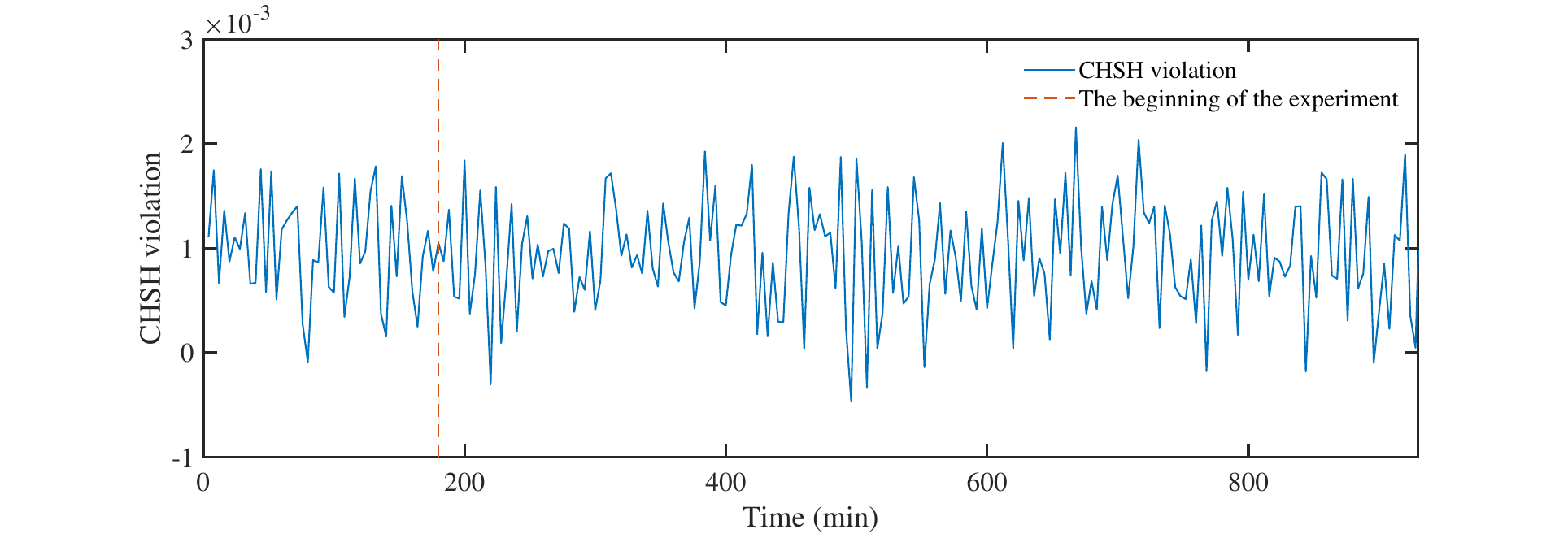}}
    \caption{CHSH violation versus time. We choose the average value of CHSH violations for each $4$~min data as a point to observe its performance over experimental time. The first 180 minutes of data corresponds to the ``training trials'', and the rest corresponds to the collected data in the main experiment of randomness expansion (we separate the two parts of data by the orange dotted line, while in the experiment we do not stop to recalibrate the system).}
    \label{Fig:chsh}
\end{figure}

During the experiment execution, we monitor the behaviour of the set-up via the real-time CHSH violation value.
Fig.~\ref{Fig:chsh} shows the measured CHSH violation value versus time. Every 180 seconds, we estimate the CHSH violation real-time with the accumulated data. The data consists of the 3-hour training and the 13.1-hour main experiment. Since the time span is short, and the CHSH violation value is relatively stable, we continuously carry out the experiment without any break.    

\subsection{Random Number Generation Result}
In our randomness extraction, we conservatively set $\varepsilon_x=2^{-100}$.
In the randomness extraction procedure, we apply a Toeplitz matrix of the size $m\times n = k_{\text{gen}}\times 2N = (5.47\times10^8)\times(2\times2.35\times10^{11})$.
The experiment succeeds in prior to the largest allowed number of trials $N$, which stops at the $N_{\text{act}}^{\text{th}}$ trial, and the raw data is of the length $2N_{\text{act}} = 2\times 1.89\times10^{11}$.
While to keep in accordance with the protocol requirement, the Toeplitz matrix is applied to the concatenation of the raw data of measurement outcomes of length $2\times 1.89\times10^{11}$ and a sequence of zeros, of which the total length is $n = 2\times2.35\times10^{11}$.
$m = 5.47\times10^8$ bits of near-uniform random bits have been extracted.

We use the fast Fourier transform (FFT) to speed up the multiplication,
\begin{equation}
	T_{m\times n}\times V_n = IFFT( FFT(T_{m+n-1}) \cdot FFT(V_m) ).
\end{equation}
Here FFT is the fast Fourier transform on the vector, $T_{m+n-1}$ is the elements $(a_{-(n-1)}, ... , a_{-1}, a_0, a_1, ..., a_{m-1})$ in the Toeplitz matrix. IFFT is the inverse fast Fourier transform of the product of the vectors. The vector dimension should be expand to $m+n-1$ by adding zeros at the end.

To save memory use in randomness extraction, for some $l\leq m,\,l\in\mathbb{N}^+$, we divide the matrix into $\lceil n/l\rceil$ blocks each with dimension $m\times l$,
\begin{equation}\label{}
  T_{m\times n} =
  \left(
    \begin{array}{cccc}
    T_{m\times l}^0 & T_{m\times l}^1 & \cdots & T_{m\times l}^{\lceil n/l\rceil-1}
    \end{array}
  \right),
\end{equation}
with block $T_{m\times l}^{i}$ given by
\begin{equation}\label{}
  T_{m\times l}^{i} =
  \left(
    \begin{array}{cccc}
    a_{-i\cdot l}     & a_{-(i\cdot l+1)} & \cdots   & a_{-(i\cdot l+l-1)}     \\
    a_{-i\cdot l+1}   & a_{-i\cdot l}     & \ddots   & a_{-(i\cdot l+l-1)+1}   \\
    \vdots            & \vdots            &          & \vdots                  \\
    a_{-i\cdot l+m-1} & a_{-i\cdot l+m-2} & \cdots   & a_{-(i\cdot l+l-1)+m-1} \\
    \end{array}
  \right).
\end{equation}
Similarly, we divide the vector into $\lceil n/l\rceil$ blocks,
\begin{equation}\label{}
  V_{n} =
  \left(
    \begin{array}{c}
    V_l^0 \\
    V_l^1 \\
    \vdots \\
    V_l^{\lceil n/l\rceil-1}
    \end{array}
  \right),
\end{equation}
with each block given by
\begin{equation}\label{}
  V_{l}^i =
  \left(
    \begin{array}{c}
    v_{i\cdot l} \\
    v_{i\cdot l+1} \\
    \vdots \\
    v_{i\cdot l+l-1} \\
    \end{array}
  \right).
\end{equation}
We then apply FFT to each block. The results are given by
\begin{equation}\label{}
  U_{m}' =
  \left(
    \begin{array}{cccc}
    U_l^0 & U_l^1 & \cdots & U_l^{\lceil n/l\rceil-1}
    \end{array}
  \right),
\end{equation}
where $U_l^i=T^i_{m\times l}\cdot V^i_l$, which is given by
\begin{equation}\label{}
  U_{l}^i =
  \left(
    \begin{array}{cc}
    u_0^i\\
    u_1^i\\
    \vdots\\
    u_{m-1}^i\\
    \end{array}
  \right).
\end{equation}
The final result is given by
\begin{equation}\label{}
  U_{m} =
  \left(
    \begin{array}{cc}
    \Sigma_i u_0^i\\
    \Sigma_i u_1^i\\
    \vdots\\
    \Sigma_i u_{m-1}^i\\
    \end{array}
  \right).
\end{equation}
The blocked algorithm is slower than the full FFT algorithm, but it saves memory.
We perform the extraction calculation on a personal computer with 32 Gbytes memory by dividing the original data into 1200 blocks (corresponding to $l=3.92\times 10^8$), which takes about 85 hrs including data loading and computation.

\subsection{Statistical analysis of output randomness}
To check the statistical properties of our output, we run it through the NIST test suite~\cite{NIST_Tests}. To do so, we set the section length to 1 Mbits for our $5.47\times10^8$ random output bits. As shown in Tab.~\ref{tab:nisttest}, the random bits successfully pass the tests.

\begin{table}[htb]
\centering
  \caption{Results of the NIST test suite dividing our output into 1 Mbit sections.}
\begin{tabular}{c|ccc}
\hline
Statistical tests & P value & Proportion & Result\\
\hline
Frequency				& 0.38288 & 0.993 & Success \\
BlockFrequency			& 0.70425 & 0.993 & Success \\
CumulativeSums			& 0.30262 & 0.994 & Success \\
Runs					& 0.77097 & 0.994 & Success \\
LongestRun				& 0.61977 & 0.991 & Success \\
Rank					& 0.95893 & 0.991 & Success \\
FFT						& 0.60438 & 0.985 & Success \\
NonOverlappingTemplate	& 0.48848 & 0.990 & Success \\
OverlappingTemplate		& 0.15201 & 0.991 & Success \\
Universal				& 0.66986 & 0.991 & Success \\
ApproximateEntropy		& 0.71183 & 0.991 & Success \\
RandomExcursions 		& 0.35459 & 0.987 & Success \\
RandomExcursionsVariant	& 0.42987 & 0.993 & Success \\
Serial 					& 0.13519 & 0.995 & Success \\
LinearComplexity		& 0.81954 & 0.987 & Success \\
\hline
\end{tabular}
\label{tab:nisttest}
\end{table}

\section{Discussion on Security Assumptions}\label{sec:localbias}
\subsection{Security Assumptions in This Work}

Since there is no unconditional randomness in nature, a certain set of assumptions or requirements is always needed for randomness generation. In Fig.~1 of the main text and the caption below it we list all the assumptions in this work. We review the assumptions here:
\begin{enumerate}
  \item Secure lab: The information exchange with an outside entity is controlled.
      The devices cannot communicate to the outside to leak the experimental results directly.
\item Non-signaling condition: In each trial, the measurement process of Alice/Bob is independent of the other party.
\item Trusted coordinator: A well characterised biased random number generator (depicted by ``Biased QRNG'' in the figure) determines a trial to be ``spot'' or ``checking''.
    The setting is private to the measurement devices and the entanglement source.
\item Trusted inputs: Alice and Bob each has a private random number generator (depicted by ``Unbiased QRNG'' in the figure) to feed perfect random bits to the measurement device in the ``checking trials''.
\item Trusted post-processors: The classical post-processing procedure is trusted.
\item Quantum mechanism: Quantum mechanics is correct and complete.
\end{enumerate}

Most of the assumptions are necessarily required in almost all device-independent quantum information processing tasks, such as DIQRNG~\cite{pironio2010,bierhorst2018experimentally,liu2018device}.
The first assumption prevents the devices from leaking the outputs of the experiment directly to an adversary, which trivially compromises security.
The second and fourth assumptions are needed so that the violation of a Bell inequality is valid and cannot be faked with classical means. In our work, we arrange the related measurement events of Alice and Bob to be space-like separated (see Sec.~\ref{spacetime}), which meets the requirement of the second assumption under the theoretical framework of relativity. The fifth assumption is needed for correct classical data processing.

One special requirement for our DIQRE experiment is the spot-checking arrangement.
We assume a trusted initial random seed to determine whether a trial is spot or checking, and the untrusted devices cannot tell whether a trial is spot or checking before the measurement is finished.
The trust on this random seed is necessary to avoid the attack where an adversary behaves differently in the checking trials and the spot trials.
Technically, this assumption does not pose additional requirements.
To see this, first, for any information processing tasks based on the loophole-free violation of Bell tests, Assumption 4 is required, which unavoidably requires trust on some initial randomness.
Apart from DIQRNG, there are other types of QRNGs, such as the source-independent QRNGs~\cite{Ma2016QRNG}, which require trusted random seed for input setting.
Besides, even QRNGs with trusted devices require initial random seed, since the randomness extraction procedure cannot be done without initial random seed.
The requirement of trusted initial random seed for randomness extraction is embedded in Assumption 5.
To realise the spot-checking protocol, the random seed for determining a trial to be spot or checking can be shared between the two valid users of DIQRE, Alice and Bob, before the experiment. They keep the seed private until it is used for the input setting. In this sense, trust on this random seed is not stronger than the Assumption 1 required for the secure store of generated randomness.

\subsection{Simulation of Randomness Expansion with Independent Local Biased Random Number Generators}
\begin{figure}[tbh]
    \centering
    \resizebox{8cm}{!}{\includegraphics{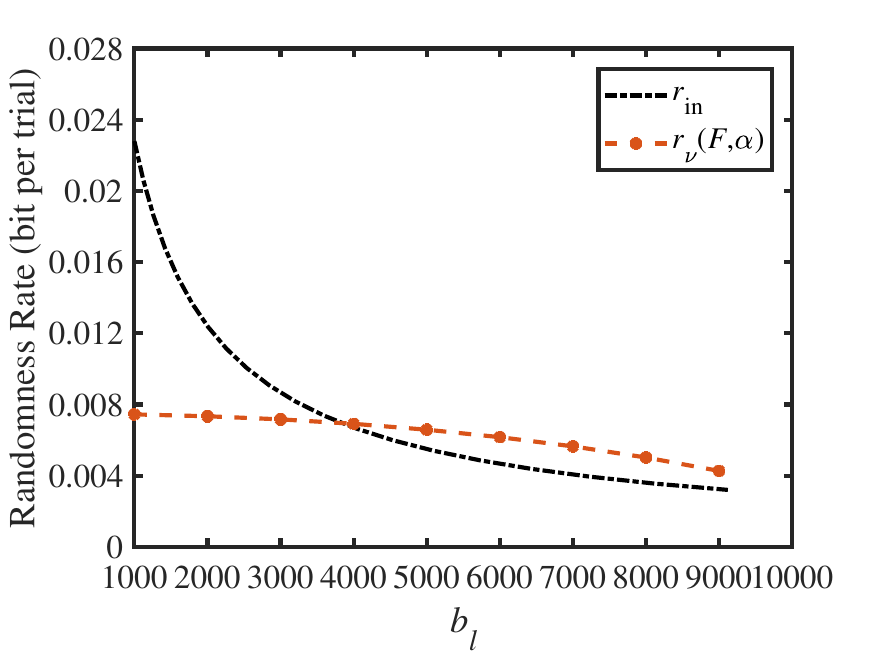}}
    \caption{The simulated expected randomness rate in the protocol with local biased random inputs (PEF). Under $\alpha=1+1.66\times10^{-9}$, we optimize the output randomness rate with different input settings (orange dash line), where $X_i,\,Y_i$ observe independent Bernoulli distributions $(1-q,q)$.
    We denote $b_l=(1-q)/q$ as the bias parameter. When $b_l$ is approximately larger than 4000, the output randomness rate exceeds the input entropy rate (black dotted line).
    }
    \label{Fig:local_bias}
    \end{figure}

In our spot-checking protocol, we rely on a trusted random number generator represented by the random variable $\bm{T} = (T_1,\cdots,T_n)$, to which an adversary is inaccessible (Assumption 3). This is a common assumption in spot-checking type of device-independent tests, see Refs.~\cite{Miller14,arnon2018practical,knill2018quantum} for example.
However, we can possibly remove it under the QPE framework and even avoid using a spot-checking protocol for DIQRE. Alice and Bob can each employ a biased random number generator to determine their input settings locally, where now the single trial experiment model is $\mathcal{M}(C,Z) = \mu_A(X)\times\mu_B(Y)\times\mathcal{M}(C|XY)$ with $\mu_A(X),\,\mu_B(Y)$ being independent Bernoulli distributions $(1-q_A,q_A),\,(1-q_B,q_B)$ with $q_A,q_B\in(0,1)$. Assuming the PEFs under this condition close to QEFs, with the power for PEFs set to $\alpha=1+1.66\times 10^{-9}$ we perform an optimization over PEFs to derive a simulated randomness expansion result as shown by Fig.~\ref{Fig:local_bias}, where we use the same input-conditional probability distribution above for simulation. Here we let $q=q_1=q_2$ and define $b_l = (1-q)/q$ as the bias parameter in this setting. We see that it is theoretically possible to realise expansion when $b_l$ is large enough. With current experimental conditions, however, it is difficult to realise randomness expansion in this way, as one may notice that the counting rate for the input setting $(1,1)$ is too low due to the highly biased local input settings, and a relatively small power in the R\'enyi power makes the implementation time far too long. Nevertheless, this provides us with a new insight on how to remove additional assumptions in a device-independent task.

\section{Comparison with Other Protocols}\label{Supp:Comparison}

\subsection{Comparison between Quantum Probability Estimation and Entropy Accumulation Theorem}
Apart from the QPE-based method, for the tasks of DIQRNG and DIQRE, there exist different theoretical approaches for analysis, for example, the results in~\cite{Vazirani12,Miller14,arnon2018practical}. Among these methods, the entropy accumulation theorem (EAT)~\cite{arnon2018practical} is the first theoretical method achieving an asymptotically optimal result for the certification of DIQRNG in the limit of infinite data size, and it has been applied to the first loophole-free DIQRNG experiment against quantum side information~\cite{liu2018device} and a recent experimental work on DIQRE~\cite{liu2019device}. Here we present a brief discussion on the pros and cons of QPE and EAT methods, and list the existing experimental works using the two methods.

(a) QPE method~\cite{knill2018quantum}: This method is designed to witness randomness in an efficient manner, especially for applications requiring a low latency for randomness generation.
A specific randomness generation target is set in advance.
Apart from this work, the QPE-based protocol has been used for DIQRNG secure against quantum side information~\cite{Zhang2018Low}, and a recent experimental work of DIQRE secure against classical side information~\cite{shalm2019device} is carried out based on a classical-proof version of the theory, the probability estimation method~\cite{zhang2018certifying}.
The QPE method takes advantage of full input-output probability distribution, and is highly numerical and sensitive in statistical fluctuations.
For example, the experimental settings in this work are optimized with respect to an empirical input-output probability distribution obtained from three hour's training data.
Besides, in all these QPE-based works, a specific randomness generation or expansion task is set, and the largest allowed number of rounds for the task is pre-determined.

(b)	EAT method~\cite{arnon2018practical}: This method utilises the violation value of a specific Bell inequality, and leads to a formula connecting the Bell test value and the amount of randomness generated depending on the number of rounds.
Using this we can calculate the expected number of rounds needed for randomness generation or expansion.
It has been applied to the first DIQRNG experiment secure against quantum side information~\cite{liu2018device} and a recent experimental work on DIQRE~\cite{liu2019device}.
In these two works, the CHSH inequality is used.
An abort condition is set in the protocol, but there is no need to set a specific amount of randomness to be generated as a target.
For example, in the work of Ref.~\cite{liu2019device}, a CHSH test is carried out to estimate the parameters of the experiment according to the value of violation, and a 19.2-hour experimental time is set to achieve expansion afterwards.

We suspect that in general, whether QPE or EAT method is the better technique to use depends on the situation.
A detailed analysis of when one is preferable to the other is desirable, but is beyond the scope of the present work.
We note that some preliminary work has been done in the Section VIII. B. of Ref.~\cite{knill2018quantum}.
From current theoretical and experimental results, we would like to state that the current QPE method is more tailored for efficiently witnessing randomness generation with a low latency than EAT-based approach.
Though our discussion is more in a qualitative manner, we hope this can benefit further investigations.

\subsection{Comparison between CHSH Bell Test and GHZ Test}
In this work, we have utilised the CHSH-type Bell test. In general, other types of Bell test can be used for randomness expansion. One specific choice is the Greenberger-Horne-Zeilinger (GHZ) protocol proposed in~\cite{Colbeck09}, where the GHZ test is used.
This is a randomness expansion protocol based on three nonlocal quantum devices. The protocol is implicitly related the Mermin inequality~\cite{mermin1990extreme}, while here we briefly state the original design.
In the randomness expansion protocol of~\cite{Colbeck09}, the randomness is generated from the outputs of the GHZ test, which goes as follows:
\begin{enumerate}
  \item The users randomly set inputs to three quantum measurement devices as $X,Y,Z\in\{0,1\}$, and the measurement devices output $A,B,C\in\{\pm1\}$. The test is repeated for a certain number of trials.
  \item The users keep the trials with the input settings $(X,Y,Z) = (0,0,0),(0,1,1),(1,0,1),(1,1,0)$.
  \item If $ABC=-1$ for the first setting and $ABC=+1$ for the remaining three, the test succeeds, and the protocol proceeds to randomness extraction.
      Otherwise the protocol aborts.
\end{enumerate}

A nice theoretical property of the GHZ test is that the condition cannot be met perfectly with classical strategies using shared randomness but with a unique quantum strategy~\cite{Colbeck09} (in the sense of local unitary operations).
Therefore, the GHZ test has a convenient cheat-detection property in the ideal case, where a user can be certain whether there is an attack by observing whether the requirement of the test is met~\cite{colbeck2011private}.
Nevertheless, in a real experiment with imperfect devices under a potentially adversarial condition, an enough amount of statistics shall be required for the GHZ protocols for the certification of the generated randomness, and tolerance to noise and error needs to be considered, too.

Compared to the GHZ protocol, we emphasize that the probabilistic nature of the CHSH test does not introduce an additional side-channel of information leakage.
In both protocols, the amount of possible information leakage can be estimated arbitrarily well over a long run of the protocol, in the sense that under the QPE framework, the soundness error of the protocol can be taken arbitrarily small by taking long enough statistics.
In general, the number of trials required to achieve a particular soundness error depends on the Bell test that is used, and it might benefit in certain conditions to use the GHZ test due to its deterministic property, i.e., the GHZ game can be won perfectly. We leave the investigation of DIQRE based on general Bell tests for further research.

\bibliographystyle{apsrev4-1}
\bibliography{BibDIQRNG_NP}

\end{document}